\documentclass[11pt]{amsart}

\usepackage{amsmath}
\usepackage{amsfonts}
\usepackage{amssymb}
\usepackage{amscd}
\usepackage{amsthm}
\usepackage{framed}
\usepackage{fullpage}
\usepackage{graphicx}
\usepackage{latexsym}
\usepackage[numbers]{natbib}  
\usepackage{multirow}
\usepackage{tikz}
\usetikzlibrary{positioning}
\usetikzlibrary{arrows}

\allowdisplaybreaks[4]

\usepackage{hyperref}
\hypersetup{colorlinks,linkcolor={red},citecolor={blue},urlcolor={blue}}

\newtheorem{theorem}{Theorem}[section]

\newtheorem{conj}[theorem]{Conjecture}

\theoremstyle{definition}

\numberwithin{equation}{section}

\newcommand{\CC}{\mathbb C}
\newcommand{\HH}{\mathbb H}

\newcommand{\NN}{\mathbb N}

\newcommand{\ZZ}{\mathbb Z}

\newcommand{\latt}[1]{{\langle{#1}\rangle}}

\def\dim{\operatorname{dim}}

\newcommand{\Eh}{E_{7+1/2}}

\newcommand{\T}{\mathsf{T}}

\usepackage{multirow}

\begin{document}

\hfill{KIAS-P23033, MPIM-Bonn-2023\\[+10mm]}

\title[On intermediate Lie algebra $E_{7+1/2}$]{On intermediate Lie algebra $E_{7+1/2}$}

\author{Kimyeong Lee}

\address{Korea Institute for Advanced Study, 85 Hoegiro, Dongdaemun-gu, Seoul 02455, Korea}

\email{klee@kias.re.kr}

\author{Kaiwen Sun}

\address{Max Planck Institute for Mathematics, Vivatsgasse 7, D-53111 Bonn, Germany}

\email{ksun@mpim-bonn.mpg.de}

\author{Haowu Wang}

\address{School of Mathematics and Statistics, Wuhan University, Wuhan 430072, Hubei, China}

\email{haowu.wangmath@whu.edu.cn}

\subjclass[2020]{}

\date{\today}


\begin{abstract}

$E_{7+1/2}$ is an intermediate Lie algebra filling a hole between $E_7$ and $E_8$ in the Deligne-Cvitanovi\'c exceptional series. It was found independently by Mathur, Muhki, Sen in the classification of 2d RCFTs via modular linear differential equations (MLDE) and by Deligne, Cohen, de Man in representation theory. In this paper we propose some new vertex operator algebras (VOA) associated with $E_{7+1/2}$ and give some useful information at small levels. We conjecture that the affine VOA $(E_{7+1/2})_k$ is rational if and only if the level $k$ is at most $5$, and provide some evidence from the viewpoint of MLDE. We propose a conjectural Weyl dimension formula for infinitely many irreducible representations of $E_{7+1/2}$, which generates almost all irreducible representations of $E_{7+1/2}$ with level $k\leq 4$. More concretely, we propose the affine VOA $E_{7+1/2}$ at level 2 and the rank-two instanton VOA associated with $E_{7+1/2}$. We compute the VOA characters and provide some coset constructions. These generalize the previous works of Kawasetsu for affine VOA $E_{7+1/2}$ at level 1 and of Arakawa--Kawasetsu at level $-5$. We then predict the conformal weights of affine VOA  $E_{7+1/2}$ at level $3,4,5$.
\end{abstract}

\maketitle

\begin{small}
\tableofcontents
\end{small}

\section{Introduction}
The \textit{Deligne-Cvitanovi\'c exceptional series} of simple Lie algebras \cite{Deligne,Cvitanovic:2008zz}
\begin{align} \label{DC}
A_1\subset A_2 \subset G_2 \subset D_4 \subset F_4 \subset E_6 \subset E_7 \subset E_8  
\end{align}
appears constantly in representation theory, number theory and theoretical physics. These simple Lie algebras $\mathfrak{g}$
exhibit many remarkable properties. For example, the decomposition of the tensor products of the adjoint representations have some uniform expressions such as
\begin{align} 
    \mathrm{Sym}^2 \mathfrak{g} &=1+Y_2+Y_2^*,
\end{align}
and the irreducible components in the decompositions have some miraculous dimension formulas (often called \textit{Deligne dimension formulas}) as rational functions of the dual Coxeter number $h^\vee$. For instance,   
\begin{align} 
    \dim \mathfrak{g} =&\, \frac{2(5h^\vee -6)(h^\vee +1)}{h^\vee +6},\\ 
\dim Y_2 =&\, \frac{5h^{\vee 2}(2h^\vee+3)(5h^\vee-6)}{(h^\vee+6)(h^\vee+12)},\\
\dim Y_2^* =&\, \frac{270 h^{\vee 2}(h^\vee+1)(h^\vee-2)}{(h^\vee+6)^2(h^\vee+12)}.
\end{align}
In 1996, Cohen and de Man \cite{Cohen} found in total 25 such dimension formulas as rational functions of $h^\vee$. Interestingly, they noticed that for $h^\vee=24$, all the dimension formulas produce integer values such as $\dim \mathfrak{g}=190$, $\dim Y_2=15504$ and $\dim Y_2^*=2640$. However, there exists no simple Lie algebra with such dual Coxeter number $24$ and dimension $190$. This exotic algebra, now known as \textit{intermediate Lie algebra $E_{7+1/2}$}, was eventually constructed by Landsberg and Manivel \cite{LM} using \textit{sextonions} which is a six dimensional algebra  lying between the quaternions $\mathbb{H}$ and octonions $\mathbb{O}$ \cite{Westbury:2004vav,Marrani:2015nta}. The name of this algebra comes from the fact that $h^\vee=24$ is intermediate between the dual Coxeter number 18 of $E_7$ and 30 of $E_8$, and that in many senses this algebra  fills the hole between $E_7$ and $E_8$ in the Deligne-Cvitanovi\'c exceptional series. The construction is inspired from the observation that the last five Lie algebras in \eqref{DC} have $h^\vee/3-2$ as $0,1,2,4,8$ which are precisely the dimensions of $0,\mathbb{R},\mathbb{C},\mathbb{H},\mathbb{O}$.

The existence of $E_{7+1/2}$ was actually discovered earlier in physics. In 1988, Mathur, Mukhi and Sen \cite{Mathur:1988na} used second order holomorphic modular linear differential equations (also known as Kaneko-Zagier equation \cite{KZ}) to classify the 2d rational conformal field theories (RCFT) with two characters (up to degeneracy), which is equivalent to classify certain $C_2$-cofinite rational VOAs. They found that there are ten possibilities in total, eight of which correspond exactly to the level 1 affine VOA $(\mathfrak{g})_1$, where $\mathfrak{g}$ belongs to the Deligne-Cvitanovi\'c exceptional series \eqref{DC}. For the remaining two possibilities, one corresponds to the Galois shuffle of the well-known Lee-Yang minimal model $M(5,2)$, while the other mysterious one has central charge $c=38/5$, non-vacuum conformal weight $h=4/5$ and dimension (spin-1 currents) $190$. It was observed in \cite{Mathur:1988na} that the well-known central charge formula for WZW model $(\mathfrak{g})_k$ 
\begin{align}
    c=\frac{k\dim(\mathfrak{g})}{k+h^\vee}
\end{align}
still holds if taking $h^\vee=24$, $\dim \mathfrak{g}=190$ and level $k=1$. 
For this reason, the last case can be regarded as the \textit{affine VOA $E_{7+1/2}$ at level 1}. Such VOA was later rigorously constructed by  Kawasetsu \cite{Kawasetsu} as an intermediate vertex subalgebra of the lattice VOA $V_{E_8}$. This suggests that the notion that $\Eh$ fills a hole between $E_7$ and $E_8$ in the Deligne-Cvitanovi\'c exceptional series can be extended to vertex algebras.

The other known VOA associated with $\Eh$ is the \textit{affine VOA $E_{7+1/2}$ at level $-5$}, which is an extension of the  quasi-lisse affine VOA $(\mathfrak{g})_{-h^\vee/6-1}$ for the Deligne-Cvitanovi\'c exceptional series studied by Arakawa and Kawasetsu \cite{AK}. These VOAs originated in the context of 4d SCFT/2d VOA correspondence proposed by 
Beem,  Lemos,  Liendo, Peelaers,  Rastelli and  van Rees \cite{Beem:2013sza}. The VOA $(\mathfrak{g})_{-h^\vee/6-1}$ for $\mathfrak{g}=A_1,A_2,D_4,E_6,E_7,E_8$ describes the Schur sector of some special 4d $\mathcal{N}=2$ SCFTs called rank-one $H_{\mathfrak{g}}$ theories \cite{Beem:2017ooy}. For example, the rank-one $H_{D_4}$ theory is just the well-known 4d $\mathcal{N}=2$ $SU(2)$ gauge theory with four fundamental matters which has superconformal symmetry and also global symmetry $D_4$. These 4d rank-one $H_{\mathfrak{g}}$ SCFTs have natural rank-$n$ generalization, whose Higgs branch of vacua is the $n$ centered $\mathfrak{g}$-instanton moduli space (see e.g. \cite{Beem:2019snk}). Therefore, these 4d $\mathcal{N}=2$ SCFTs are also called the instanton SCFTs. We follow the name to call these VOAs of negative levels with (possible) 4d SCFT origin as \textit{instanton VOAs}. The vacuum character of these VOAs are expected to coincide with the Schur indices of the corresponding 4d SCFTs (if they exist), and the associated varieties of the VOAs are expected to coincide with the Higgs branches of the SCFTs \cite{Beem:2017ooy}, which for the Class S type are also the Moore-Tachikawa symplectic varieties \cite{Moore:2011ee}. See more discussions on the Class S type in \cite{Beem:2014rza,Arakawa:2018egx}. 

Two natural questions arise. First, do there exist more VOAs associated with intermediate Lie algebra $E_{7+1/2}$ besides the two known ones at levels $1$ and $-5$? Second, can one find more irreducible representations of $E_{7+1/2}$ besides those found by Cohen-de Man \cite{Cohen} and Landsberg-Manivel \cite{LM}?
We give affirmative answers to both questions in this paper. For the first question, there are two directions of generalizations: from affine VOA at level $1$ to higher level $k$, and from rank-one instanton VOA to higher rank-$n$. Recently, the (conjectural) characters of affine VOA $(\Eh)_2$ were successfully obtained by a Hecke operation in \cite[Section 7.1]{Duan:2022ltz}. These characters provide very useful information for our study. Though we still could not compute the characters of affine VOA $(\Eh)_k$ for $k>2$, we can predict all conformal weights and rationality of these VOAs. For the rank-$2$ VOA, we can compute  its vacuum character from MLDE,  and also for the rank-$3$, its vacuum character to a certain extent.

The basic assumption we make is that there is a one-to-one correspondence between the $\Eh$ irreducible representations and the $E_7$ irreducible representations. In other words, $\Eh$ irreducible representations are marked by $E_7$ Dynkin labels. This assumption is true for level 1 and 2 representations, as at these two levels, the affine characters have been constructed by Hecke operators or MLDEs. Under this assumption, we predict the quadratic Casimir invariants of all  $\Eh$ irreducible representations (see \eqref{eq:C2}) and the dimensions of infinitely many of them (see \S \ref{sec:weyl}). The latter is owing to certain generalizations of the familiar Weyl dimension formula.

Some of our main results can be summarized as follows:
\begin{conj}\label{conj1}
There exists rational affine VOA $(\Eh)_k$ at level $k$ if and only if $1\le k\le 5$. For these levels, $(\Eh)_k$ has central charge $190k/(24+k)$, and the number $r(k)$ of its characters coincides with that of the affine VOA $(E_7)_k$, and the characters satisfy a MLDE with the same index $l(k)$ as the one for $(E_7)_k$. Moreover, the characters of VOA $(\Eh)_2$ are defined in \eqref{eq:chilevel2}, and the conformal weights of VOAs $(\Eh)_3$, $(\Eh)_4$ and $(\Eh)_5$ are formulated in \eqref{eq:weights3}, \eqref{eq:weights4} and \eqref{eq:weights5}, respectively. The values of $r(k)$ and $l(k)$ are as follows
\begin{table}[h]
  \centering
  \begin{tabular}{c|c|c|c|c|c}\hline
    $k$ & $1$ & $2$ & $3$ & $4$ & $5$    \\ \hline
$r(k)$  & $2$ & $6$ & $12$ & $25$  & $ 44$   \\ \hline
$l(k)$ & $0$ & $0$ & $20$ & $160$ & $636$    \\ \hline
     \end{tabular}
\end{table}
\end{conj}
The irrationality of VOA $(\Eh)_k$ for $k>5$ (if they exist) was pointed out to us by Tomoyuki Arakawa and Kazuya Kawasetsu from the viewpoint of $W$-algebras. We will provide new evidence from the viewpoint of MLDE in Section \ref{sec:irr}. 

Coset constructions are very useful in the study of rational VOAs. We find two coset constructions involving affine VOA $(\Eh)_2$ which we summarize as follows.
\begin{conj}
The following isomorphisms between the coset VOAs and the Virasoro minimal models hold:
\begin{equation}
\frac{(E_{8})_2}{(\Eh)_2}= M_{\rm eff}(13,4)\quad\textrm{and}\quad \frac{(\Eh)_1\otimes (\Eh)_1}{(\Eh)_2}=M_{(D_6,A_{12})}(13,10).
\end{equation}
Here $M_{\rm eff}(13,4)$ denotes the effective description of minimal model $M(13,4)$ with effective central charge $23/26$, and $M_{(D_6,A_{12})}(13,10)$ denotes the $(D_6,A_{12})$-type non-diagonal modular invariant of minimal model $M(13,10)$.
\end{conj}
\noindent
The notion of the effective description of minimal models and the non-diagonal modular invariants will be elaborated later. These new cosets may remind one of the familiar maverick cosets found in 90s \cite{CFT} such as
\begin{equation}
\frac{(E_{8})_2}{(E_{7})_2}= (A_1)_2\otimes M(5,4)\quad\textrm{and}\quad \frac{(E_7)_1\otimes (E_7)_1}{(E_7)_2}=M(5,4).
\end{equation}

It is worth mentioning that the concept of intermediate Lie algebras is not new. In fact, the intermediate Lie algebras $A_{n-1/2},$ $B_{n-1/2},$ $C_{n-1/2},$ $D_{n-1/2}$ and their highest-weight representations have been extensively studied by Shtepin for the last three decades \cite{ShtepinAC,ShtepinB,ShtepinD0,ShtepinD}. A characteristic feature of these intermediate Lie algebras is the multiplicity-free filtration. Shtepin also found some Weyl-type character formulas and Weyl-type dimension formulas for the representations of these intermediate Lie algebras. The intermediate Lie algebra  $\Eh$ seems to be more difficult than these classical types, in the meantime more extraordinary as it fills the hole in the Deligne-Cvitanovi\'c exceptional series, thus share many similar remarkable  properties like the Deligne dimension formulas.

Besides, $\Eh$ is closely related to $W$-algebras. For example, the characters of affine VOA $(\Eh)_1$ coincide with the Ramond twisted irreducible characters of $W$-algebras $W_{-5}(E_8,f_{\theta})$ where $f_{\theta}$ is the minimal nilpotent element associated with the highest root $\theta$ \cite{Kawasetsu18}. Such a nice coincidence might persist to higher levels. Moreover, $\Eh$ seems to be also related to a 12 dimensional self-duality equation found by Devchand \cite{Devchand:2012xs}. We expect to see more applications of $\Eh$ in the future. 

\section{Known results about $\Eh$}
\subsection{Affine VOA $\Eh$ at level $1$} 
The affine VOA $(\Eh)_1$ was constructed as an intermediate vertex subalgebra of lattice VOA $V_{E_8}$ by Kawasetsu \cite{Kawasetsu}. We review the known features of this VOA here. Historically,  
for level 1 Deligne-Cvitanovi\'c exceptional series \eqref{DC}, there are two characters $\chi_0$ and $\chi_h$ as the solution of second order MLDE found in the Mathur-Muhki-Sen classification \cite{Mathur:1988na}, see also \cite{Kaneko:2013uga} for the rigorous classification. The second order holomorphic MLDE, i.e., the Kaneko-Zagier equation \cite{KZ} is
\begin{equation}\label{eq:level1mlde}
[D^{2}+\lambda E_4]\chi=0.    
\end{equation}
The central charge, the non-vacuum conformal weight, and the coefficient in the MLDE can be expressed uniformly by the dual Coxeter number $h^\vee$ as
\begin{equation}
 c=\frac{2 (5 h^\vee - 6) }{(h^\vee + 6)},\quad\quad h=\frac{h^\vee}{h^\vee+6} , \quad\textrm{and}\quad \lambda = \frac{(6-5h^\vee)(6+7h^\vee)}{144(6+h^\vee)^2} .
\end{equation}

With the dual Coxeter number $h^\vee=24$ for $E_{7+1/2}$, the  second order MLDE for the characters of affine VOA $(\Eh)_1$ becomes
\begin{equation}
\Big[D^{2}-\frac{551}{3600} E_4\Big]\chi=0.    
\end{equation}
The two solutions of the MLDE have the following Fourier expansions
\begin{align}
\chi_0&=q^{-\frac{19}{60}}(1+190 q+2831 q^2+22306 q^3+129276 q^4+611724 q^5+2511667 q^6+\dots),\\ \label{eq:chi45}
\chi_{\frac45}&=q^{\frac{29}{60}}(57+ 1102 q + 9367 q^2 + 57362 q^3 + 280459 q^4 + 1181838 q^5 + 
 4435740 q^6  +\dots) .
\end{align}
They compute the graded dimensions of VOA $(\Eh)_1$ and its intermediate module \cite{Kawasetsu}. A nice coset construction of VOA $(\Eh)_1$ proved in \cite{Kawasetsu} is
\begin{equation}\label{cosetlevel1a}
(\Eh)_1=\frac{(E_7)_1}{M(5,3)}=(E_7)_1\otimes M_{\rm eff}(5,3).    
\end{equation}
Here $M_{\rm eff}(5,3)$ denotes the effective description of minimal model $M(5,3)$, which has effective central charge $3/5$.\footnote{The effective description $M_{\rm eff}(p,q)$ of non-unitary minimal model $M(p,q)$, i.e., $p-q>1$, makes a shuffle among the vacuum and non-vacuum conformal primaries, while keep the characters unchanged. The {effective central charge} and {effective conformal weights} are given by
\begin{align}
    c_\text{eff} = 1 - \frac{6}{pq},\ \textrm{ and }\ h^{\text{eff}}_{r,s} = \frac{(pr -qs)^2 - 1}{4pq},\quad 1\leq r < q, \  1\leq s < p, \  pr > qs.
\end{align} } 
The character relations of this coset were given in \cite[Equation 7,8]{Kawasetsu}, which allows one to obtain the \textit{flavored characters}, i.e., to express the Fourier coefficients of the character $q$-series as linear combinations of $E_7$ irreducible representations. Benefited from our new results, we are able to express the flavored characters of VOA $(\Eh)_1$ by $\Eh$ irreducible representations in  
\eqref{eq:chilevel1a} and \eqref{eq:chilevel1b}.

Another simple coset involving VOA $(\Eh)_1$ is 
\begin{equation}\label{cosetlevel1b}
(\Eh)_1=\frac{(E_8)_1}{(LY)_1}.    
\end{equation}
Here $(LY)_1$ is the Lee-Yang model at level $1$, which is the effective  minimal model $M_{\rm eff}(5,2)$. It has central charge $2/5$ and non-vacuum conformal weight $1/5$, contrast to the original minimal model $M(5,2)$ which has central charge $-22/5$ and non-vacuum conformal weight $-1/5$. 
The character relation of coset \eqref{cosetlevel1b} is simply
\begin{align}
    \chi_0^{(\Eh)_1}\chi_0^{(LY)_1} +\chi_{\frac45}^{(\Eh)_1}\chi_{\frac15}^{(LY)_1} =\chi_0^{(E_8)_1}. 
\end{align}
The above character relation can be refined by turning on the $\mathfrak{g}$ fugacities, i.e., extending modular forms to Jacobi forms. 
Naively, one might suspect the left hand side has one less elliptic fugacities than the right hand side. 
However, this can be remedied by refining the $(LY)_1$ characters to the supercharacter of Lie superalgebra $B_{0,1}$ (i.e., $osp(1\vert 2)$, see e.g. \cite{creutzig2018representation}) at level 1, which can also regarded as the affine characters of intermediate Lie algebra $A_{1/2}$ at level $1$.

Recently, the two characters of $(\Eh)_1$ were realized by Harvey and Wu as the $\T_{19}$ Hecke image of the Lee-Yang model \cite[Section 5.2]{Harvey:2018rdc}. One consequence of the Hecke relation is that each character of $(\Eh)_1$ can be written as a degree 19 homogeneous polynomial of the two Lee-Yang characters \cite{Kaneko:2013uga}:
\begin{align}   \chi_0&=\phi_1^{19}+171\phi_1^{14}\phi_2^5+247\phi_1^9\phi_2^{10}-57\phi_1^4\phi_2^{15},\\
\chi_{\frac45}&=\phi_2^{19}-171\phi_2^{14}\phi_1^5+247\phi_2^9\phi_1^{10}+57\phi_2^4\phi_1^{15},
\end{align}
where $\phi_1,\phi_2$ are the Roger--Ramanujan functions 
\begin{align} 
\phi_1=q^{-\frac{1}{60}}\prod_{n=0}^{\infty}\frac{1}{(1-q^{5n+1})(1-q^{5n+4})},\qquad
\phi_2=q^{\frac{11}{60}}\prod_{n=0}^{\infty}\frac{1}{(1-q^{5n+2})(1-q^{5n+3})}.
\end{align}

From the leading Fourier coefficients of the non-vacuum character \eqref{eq:chi45}, we can recognize the fundamental representation of $\Eh$ is $\mathbf{57}$. The fundamental and adjoint representations of $\Eh$ decompose under the embedding $E_7\subset \Eh$ as
\begin{align}\label{eq:level1decomp}
\bf 57&=\bf 56+1 ,\\ \label{eq:level1decomp2}
\bf 190  &= \bf 133+56+1.
\end{align}
It is easy to check both sides have the same indices (defined in \eqref{eq:defindex}) as $6$ and $24$ respectively.

\subsection{Affine VOA $\Eh$ at level $-5$}\label{sec:rank1}
The affine VOA $(\Eh)_{-5}$ is the analogy of one-instanton VOA $({\mathfrak{g}})_{-{h^\vee}/{6}-1}$ for the Deligne-Cvitanovi\'c  exceptional series. These non-unitary and irrational VOAs come from the study on 4d SCFT/VOA correspondence \cite{Beem:2013sza,Beem:2017ooy,Beem:2019tfp}. They have negative central charge
$  c=-2-2h^\vee$. The vacuum characters of these VOAs satisfy a different uniform second order modular linear differential equation \cite{AK}
  \begin{equation}\label{eq:MLDE2b}
\Big[D^2-\frac{(h^\vee+1)(h^\vee-1)}{144}E_4\Big]\chi_{vac}=0.
\end{equation}
Notice the coefficient of $E_4$ here is different from the one for the level $1$ series discussed in the last subsection. The MLDE \eqref{eq:MLDE2b} does not have the second rational solution when $h^\vee$ is a multiple of 6. Instead it has a Log solution besides the vacuum character, see e.g. \cite[Appendix C]{Beem:2017ooy}. The associated variety of VOA $({\mathfrak{g}})_{-{h^\vee}/{6}-1}$ for the Deligne-Cvitanovi\'c  exceptional series has been studied in \cite{AM}.

When $h^\vee=24$, the vacuum character of the VOA $(\Eh)_{-5}$ is expected to satisfy the second order MLDE
\begin{equation}\label{eq:E7half-5}
\Big[D^2-\frac{575}{144}E_4\Big]\chi_{vac}=0.
\end{equation}
It is easy to solve the vacuum character as
\begin{equation}
\chi_{vac}=q^{\frac{25}{12}}(1 + 190 q + 15695 q^2 + 783010 q^3 + 27319455 q^4 + 725679750 q^5+\dots).
\end{equation}
As shown in \cite{KK,Kaneko:2013uga}, this is a quasi-modular form with exact expression 
\begin{equation}
\chi_{vac}=\frac{E_4'}{240\eta^{10}}P_3\Big(\frac{E_6}{\Delta^{1/2}}\Big)-\eta^2Q_3\Big(\frac{E_6}{\Delta^{1/2}}\Big),
\end{equation}
where $P_3(x)=x^3+904x$, $Q_3(x)=x^2+442$ \cite[Theorem 2]{KK}. Benefited from the study on irreducible representations of $\Eh$ which will be discussed later, 
we are able to determine the flavored vacuum character (up to the prefactor) as
\begin{align}\label{eq:level-5}
&1 + \mathbf{190} q + (\mathbf{15504+190+1}) q^2 + (\mathbf{749360} +2\cdot \mathbf{15504}+\mathbf{2640} +2  ) q^3\\ \nonumber
&+(\mathbf{24732110+1770496+749360+17765}+3\cdot\mathbf{15504}+\mathbf{2640}+3\cdot\mathbf{190}+2)q^4+\dots.
\end{align}
We observe that all representations appearing here are \textit{bosonic}, a term which will be defined in Section \ref{sec:weyl}. 

Interestingly, for $\mathfrak{g}=D_4,E_6,E_7$, the VOA $({\mathfrak{g}})_{-{h^\vee}/{6}-1}$ was recently suggested to connect with the curved $\beta\gamma$  system on the cone over the complex Grassmannian
$\mathrm{Gr}(2,4)$, the complex orthogonal Grassmannian
$\mathrm{OG}^+(5,10)$ and the complex Cayley plane $\mathbb{OP}^2$ respectively \cite{Eager:2019zrc}. In particular, the $E_6$ case is related to the pure spinor formulation of 10 dimensional superstring theory. It is intriguing to consider whether the VOA $(\Eh)_{-5}$ has analogous connections. 

\section{Affine VOA $\Eh$ at level 2}\label{sec:level2}
How can we define affine VOA $(\Eh)_{k}$ for $k>1$?  Assuming the general formulas for the central charge and conformal weights of the affine VOA generated by a simple Lie algebra still hold, then for affine VOA $(\Eh)_{k}$, we have
    \begin{align}\label{eq:conjch}
 c_k=\frac{{190}k}{{24}+k},\quad\text{and}\quad h_\lambda=\frac{C_2(R_\lambda)}{2({24}+k)}. 
 \end{align}
Here $C_2(R_\lambda)$ is an analogy of quadratic Casimir invariant for simple Lie algebras. Different from the case of simple Lie algebras, $C_2(R_\lambda)$ is no longer defined by $\langle \lambda+2\rho,\lambda\rangle$ in general, even if one introduces a modified Weyl vector $\rho$ for $\Eh$, e.g. \eqref{eq:weylvector}. 

Nevertheless, we can still determine the quadratic Casimir invariant for some representations of $\Eh$ from VOA characters. For example, from the non-vacuum conformal weight $\frac45$ of  
VOA $(\Eh)_1$, we derive $C_2(\mathbf{57})=40$. 
Recall the index $I$ of a representation $R$ is related to the quadratic Casimir invariant by
\begin{align}\label{eq:defindex}
    I(R)=\frac{ |R|\times C_2(R)}{2\dim(\mathfrak{g})}.
\end{align}
See e.g. textbook \cite[p 512]{CFT}. It follows that $I(\mathbf{57})=6$. For the adjoint representation, we have that $I(\mathbf{190})=h^\vee=24$ and thus $C_2(\mathbf{190})=48$.

The characters of affine VOA $\Eh$ at level 2 were recently realized as the $\T_{19}$ Hecke image of minimal model $M(13,2)$ \cite[Section 7.1]{Duan:2022ltz}. We record the Fourier coefficients in the following:
\begin{equation}\label{eq:chilevel2}
    \begin{aligned}
\chi_0=\,&q^{-\frac{95}{156}} (1 + 190 q + 18335 q^2 + 448210 q^3 + 6264585 q^4 + 62455698 q^5 +\dots),\\
\chi_{\frac{10}{13}}=\,&q^{\frac{25}{156}} (57 + 10830 q + 321575 q^2 + 4979330 q^3 + 53025295 q^4+\dots),\\
\chi_{\frac{12}{13}}=\,&q^{\frac{49}{156}} (190 + 20596 q + 537890 q^2 + 7761500 q^3 + 79066030 q^4+\dots),\\
\chi_{\frac{18}{13}}=\,&q^{\frac{121}{156}} (1045 + 48070 q + 910955 q^2 + 10983690 q^3 + 99272435 q^4+\dots),\\
\chi_{\frac{19}{13}}=\,&q^{\frac{133}{156}} (2640 + 109155 q + 1979610 q^2 + 23245740 q^3 + 206319480 q^4+\dots),\\
\chi_{\frac{21}{13}}=\,&q^{\frac{157}{156}} (1520 + 51395 q + 860890 q^2 + 9606457 q^3 + 82347710 q^4+\dots).
    \end{aligned}
\end{equation}
There are several strong pieces of evidence that these are indeed the characters of $(\Eh)_2$. We summarize as follows.
\begin{enumerate}
 \item All Fourier coefficients are positive integers. This can be proved from the definition of Hecke operators and the property of minimal model characters. 
\item  The spin-1 currents, i.e., the subleading Fourier coefficient of the vacuum character $\chi_0$ is 190, which equals the dimension of $\Eh$.  
\item All initial Fourier coefficients appear as dimensions of irreducible representations predicted by Cohen and de Man in \cite{Cohen}. Indeed, $1045,2640,1520$ are the dimensions of representations $-Y_4^*,Y_2^*,Y_3^*$ in the notion of \cite{Cohen} respectively. 
\item All six primaries have the correct conformal weights. In fact, the quadratic Casimir invariants of $\bf1045,2640,1520$  can be computed as $ 72,76,84$ respectively by the following \emph{unique} decomposition as $E_7$ irreducible representations:
\begin{align}\label{eq:level2decomp}
\bf    1045 &=\bf 912+133, \\ \label{eq:level2decomp2}
\bf 2640&=\bf 1539+912+133+56, \\ \label{eq:level2decomp3}
\bf 1520&=\bf 1463+56+1.
\end{align}
As the embedding $E_7\subset \Eh$ is maximal, the two sides of a  representation decomposition should have the same indices. Then from the known indices of $E_7$ representations, one can compute the indices of the above $\Eh$ representations, and then the quadratic Casimir invariants by \eqref{eq:defindex}. 
\end{enumerate}

Moreover, all six (or less) conformal weights of affine VOA $(\mathfrak{g})_2$ for $\mathfrak{g}$ belonging to the Deligne-Cvitanovi\'c exceptional series \eqref{DC} including $\Eh$ can be obtained from some simple rational functions of $h^\vee$.  We collect the results in \eqref{DC2cw} in Appendix I.

It was found in \cite[Section 7.1]{Duan:2022ltz} that the six characters in \eqref{eq:chilevel2} satisfy a 6th order MLDE:
\begin{equation}
    \begin{aligned}
&[{ D}^{6}+\mu_1E_4{ D}^{4}+\mu_2E_6{ D}^{3}+\mu_3E_4^2{D}^{2}+\mu_4E_4E_6{ D}+(\mu_5E_4^3+\mu_6E_6^2)]\chi=0,\\
&\ \ \mu_1=-\frac{1225}{1872},\quad \mu_2=\frac{25205}{36504},\quad \mu_3=-\frac{1349885}{3504384} ,\quad \mu_4= \frac{36703535}{296120448} \\
&\ \ \mu_5= -\frac{57214927525}{4804258148352},\quad\mu_6=-\frac{3824637775}{450399201408}.
\end{aligned}
\end{equation}
As a byproduct, we find the uniform 6th order holomorphic MLDE satisfied by the level 2 affine characters associated with the Deligne-Cvitanovi\'c exceptional series \eqref{DC} including $\Eh$. All coefficients of the MLDE are written as some rational functions of the dual Coxeter number.  These results are collected in Appendix I.

We newly find two coset constructions involving VOA $(\Eh)_2$. Firstly, we have\footnote{This coset is also found in an unpublished draft of Arakawa, Creutzig and Kawasetsu in a more general context. We thank Creutzig for pointing out to us.}
\begin{equation}
\frac{(E_{8})_2}{(\Eh)_2}= M_{\rm eff}(13,4).
\end{equation}
Both sides have central charge $\frac{23}{26}$. Denote $\mathfrak{g}=\Eh$, 
we find the following precise character relations. Since $M(13,4)$ and $ M_{\rm eff}(13,4)$ have the same characters, here we use the $M(13,4)$ primary labels:
\begin{equation}
    \begin{aligned}
\chi^{(E_8)_2}_{0} =&\,\chi_{1,3}\chi_{0}^{(\mathfrak{g})_2 } +\chi_{1,5}\chi_{\frac{10}{13}}^{(\mathfrak{g})_2 }+\chi_{1,7}\chi_{\frac{12}{13}}^{(\mathfrak{g})_2 }+\chi_{1,11}\chi_{\frac{18}{13}}^{(\mathfrak{g})_2 } +\chi_{1,9}\chi_{\frac{19}{13}}^{(\mathfrak{g})_2 }+\chi_{1,1}\chi_{\frac{21}{13}}^{(\mathfrak{g})_2 } , \\
\chi^{(E_8)_2}_{\frac{15}{16}}=&\, \chi_{2,3}\chi_{0}^{(\mathfrak{g})_2 } +\chi_{2,5}\chi_{\frac{10}{13}}^{(\mathfrak{g})_2 }+\chi_{2,7}\chi_{\frac{12}{13}}^{(\mathfrak{g})_2 }+\chi_{2,11}\chi_{\frac{18}{13}}^{(\mathfrak{g})_2 } +\chi_{2,9}\chi_{\frac{19}{13}}^{(\mathfrak{g})_2 }+\chi_{2,1}\chi_{\frac{21}{13}}^{(\mathfrak{g})_2 }  ,\\
\chi^{(E_8)_2}_{\frac32}=&\, \chi_{3,3}\chi_{0}^{(\mathfrak{g})_2 } +\chi_{3,5}\chi_{\frac{10}{13}}^{(\mathfrak{g})_2 }+\chi_{3,7}\chi_{\frac{12}{13}}^{(\mathfrak{g})_2 }+\chi_{3,11}\chi_{\frac{18}{13}}^{(\mathfrak{g})_2 } +\chi_{3,9}\chi_{\frac{19}{13}}^{(\mathfrak{g})_2 }+\chi_{3,1}\chi_{\frac{21}{13}}^{(\mathfrak{g})_2 }    .
\end{aligned}
\end{equation}
The three characters of $(E_8)_2$ correspond to $\bf 1,248,3875$ irreducible representations respectively. 

The second coset construction is more intricate. 
Consider the following block-diagonal modular invariant of $M(13,10)$. We choose the following extended primaries
\begin{align}
   \chi_{1,i}&=\chi^{M(13,10)}_{1,i} +\chi^{M(13,10)}_{1,13-i},\quad i=1,2,\dots,6,\textrm{ with weights }0,\frac{1}{13},\frac{7}{13},\frac{18}{13},\frac{34}{13},\frac{55}{13},\\
    \chi_{3,i}&=\chi^{M(13,10)}_{3,i} +\chi^{M(13,10)}_{3,13-i},\quad i=1,2,\dots,6,\textrm{ with weights }\frac{8}{5},\frac{44}{65},\frac{9}{65},-\frac{1}{65},\frac{14}{65},\frac{54}{65},\\   \chi_{5,i}&=\chi^{M(13,10)}_{5,i},\qquad\qquad\qquad i=1,2,\dots,6,\textrm{ with weights }  \frac{29}{5},\frac{252}{65},\frac{152}{65},\frac{77}{65},\frac{27}{65},\frac{2}{65}.
\end{align}
They form the $(D_6,A_{12})$-type modular invariant of $M(13,10)$:
\begin{equation}
    Z_{(D_6,A_{12})}=\sum_{i=1}^6(|\chi_{1,i}|^2+|\chi_{3,i}|^2+2| \chi_{5,i}|^2).
\end{equation}
The classification of modular invariants of minimal models is a classical result (see e.g. \cite{CFT}). 
We then propose the coset
\begin{equation}
\frac{(\Eh)_1\otimes (\Eh)_1}{(\Eh)_2}=M_{(D_6,A_{12})}(13,10).
\end{equation}
Both sides have central charge $\frac{38}{65}$.\footnote{
The coset on the left hand side has the same central charge as $M(13,10)$ was recently noticed in  \cite[p 56]{Cheng:2020srs}.} 
We find the following precise character relations
\begin{align}
  \chi^{(\mathfrak{g})_1}_0\otimes \chi^{(\mathfrak{g})_1}_{0}=&\,\chi_{1,1}\chi_{0}^{(\mathfrak{g})_2 } -\chi_{1,6}\chi_{\frac{10}{13}}^{(\mathfrak{g})_2 }+\chi_{1,2}\chi_{\frac{12}{13}}^{(\mathfrak{g})_2 }-\chi_{1,5}\chi_{\frac{18}{13}}^{(\mathfrak{g})_2 } +\chi_{1,3}\chi_{\frac{19}{13}}^{(\mathfrak{g})_2 }+\chi_{1,4}\chi_{\frac{21}{13}}^{(\mathfrak{g})_2 } , \\
\chi^{(\mathfrak{g})_1}_0\otimes \chi^{(\mathfrak{g})_1}_{\frac45}=&\, -\chi_{5,1}\chi_{0}^{(\mathfrak{g})_2 } +\chi_{5,6}\chi_{\frac{10}{13}}^{(\mathfrak{g})_2 }-\chi_{5,2}\chi_{\frac{12}{13}}^{(\mathfrak{g})_2 }+\chi_{5,5}\chi_{\frac{18}{13}}^{(\mathfrak{g})_2 } -\chi_{5,3}\chi_{\frac{19}{13}}^{(\mathfrak{g})_2 }-\chi_{5,4}\chi_{\frac{21}{13}}^{(\mathfrak{g})_2 }  ,\\
\chi^{(\mathfrak{g})_1}_{\frac45}\otimes \chi^{(\mathfrak{g})_1}_{\frac45}=&\, \chi_{3,1}\chi_{0}^{(\mathfrak{g})_2 } -\chi_{3,6}\chi_{\frac{10}{13}}^{(\mathfrak{g})_2 }+\chi_{3,2}\chi_{\frac{12}{13}}^{(\mathfrak{g})_2 }-\chi_{3,5}\chi_{\frac{18}{13}}^{(\mathfrak{g})_2 } +\chi_{3,3}\chi_{\frac{19}{13}}^{(\mathfrak{g})_2 }+\chi_{3,4}\chi_{\frac{21}{13}}^{(\mathfrak{g})_2 }    .  
\end{align}
Note the degeneracy 2 is consistent on the two sides of the coset relation. The negative signs in the above character relations reflect the novelty of intermediate Lie algebra. 

We remark that unlike the level 1 case, the two cosets we find for $(\Eh)_2$ do not give us explicit formulas for its six characters. It would be ideal if one can express $(\Eh)_2$ characters by familiar VOA characters, for example, by realizing the coset $(\Eh)_2/(E_7)_2$ as a product of some minimal models and WZW models. Besides, the $(\Eh)_1$ characters are known to have a certain Nahm sum expression \cite[Equation 9,10]{Kawasetsu}. It would be interesting to find whether $(\Eh)_2$ characters allow similar expressions.

\section{Affine VOA $\Eh$ at higher levels}
An imminent question is how to define affine VOA $(\Eh)_k$ for  level $k>2$. Albeit numerous trying, currently we do not know how to define it for level $3$ or compute the characters. The Hecke operator approach unfortunately ceases to work for level $k>2$. Nevertheless, we can still make some reasonable speculations on the general structure of affine VOA $(\Eh)_k$. Our main inspiration comes from the similarity between affine VOAs $(\Eh)_k$ and $(E_7)_k$ for $k=1,2$. 

It is well-known the number $r(k)$ of characters of affine VOA $(E_7)_k$ are generated by the series
\begin{align}
\sum r(t)x^t&=\frac{1}{(1-x)^2(1-x^2)^3(1-x^3)^2(1-x^4)}\\
&=1+2 x+6 x^2+12 x^3+25 x^4+44 x^5+79 x^6+O(x^{7}).\nonumber
\end{align}

The characters of affine VOA $(E_7)_k$ have good modularity, i.e., they form a weakly holomorphic vector-valued modular form of degree $r(k)$ and weight zero. In the meantime, they must satisfy a MLDE of degree $r(k)$ with a non-negative integer index $l(k)$ which counts the number of poles of the coefficients $\phi_i$ below. 

The general form of a degree $d$ MLDE is
\begin{align}
    \Big[D^d+\sum_{i=0}^{d-1}\phi_i(\tau)D^{i}\Big]\chi(\tau)=0,
\end{align}
where $\phi_i$ is a meromorphic modular form of weight  $2d-2i$ on $\mathrm{SL}(2,\ZZ)$. Assume there are in total $d$ number of independent solutions in the form $q^{\alpha_i}\sum_{j=0}^\infty a_{ij}q^j$, $q=e^{2\pi i\tau}$. 
Then the theory of MLDE yields that the index $l$ is related to the degree $d$ and the exponents $\alpha_i$ of the solutions by a Wronskian analysis \cite{Mathur:1988na}
\begin{align}\label{eq:ldef}
\frac{l}{6}=\frac{d(d-1)}{12}-\sum_{i=1}^{d}\alpha_i,\ \textrm{   where   }\ \alpha_i=-\frac{c}{24}+h_i.
\end{align}
See more review in e.g. \cite[Section 2.2]{Duan:2022ltz}. It turns out that the index $l$ of MLDE is very useful information to study the structure of VOAs $(\Eh)_k$, which eventually leads to our main Conjecture \ref{conj1}. To obtain such information, we first introduce the Weyl vector for $\Eh$ and a conjectural formula of the quadratic Casimir invariants for the irreducible representations of $\Eh$.

\subsection{Weyl vector}
Define the Weyl vector for $\Eh$ as the sum of all positive roots, i.e.,
\begin{equation}\label{eq:weylvector}
\rho=\rho_{E_7}+\rho_{\bf 56}=\frac{1}{2}\Big(\Delta_++\frac{1}{2}\mathbf{56}\Big).
\end{equation}
The half weights of $\bf 56$ are chosen to have positive intersection numbers with $\rho_{E_7}$. Explicitly, we have in the fundamental weight basis
\begin{equation}
\rho_{E_7}=(1,1,1,1,1,1,1),\qquad \rho_{\bf 56} =(0,0,0,1,0,1,1).
\end{equation}
The $E_7$ Dynkin diagram with node ordering is in Figure \ref{fig:E7dynkin}. 
It is easy to find that for $\mathfrak{g}=\Eh$ and $h^\vee=24$ we have
\begin{align}
    \langle\rho,\rho\rangle_{E_7}=\frac{h^\vee\dim(\mathfrak{g})}{12}=380.
\end{align}
This means that $\Eh$ satisfies the \textit{Freudenthal-de Vries strange formula} just like simple Lie algebras.

The fundamental weights $w_4,w_6,w_7$ (recall $\rho_{\bf 56}=w_4+w_6+w_7$) are special. In some sense they are like fermions, i.e. with odd nature, thus we call them \textit{fermionic fundamental weights}. We draw the Dynkin diagram of $E_7$ and $\Eh$ and mark the irreducible representations associated with the fundamental weights in Figures \ref{fig:E7dynkin} and \ref{fig:E75dynkin}. Luckily the dimensions of all these seven irreducible representations of $\Eh$ can be deduced from the dimension formulas in \cite{Cohen}.

Notice that $\rho=w_1+w_2+w_3+2w_4+w_5+2w_6+2w_7$. We modify the comarks of $w_4$, $w_6$ and $w_7$ from $3$, $1$ and $2$ to $6$, $2$ and $4$, respectively. 
In this case,  $\Eh$ has comarks
$$a^\vee=(1,2,3,4,6,2,2,4),$$
and the summation of comarks gives the dual Coxeter number $h^\vee=24$. 

\begin{figure}[h]
 \caption{Dynkin diagram of $E_7$ and irreducible representations associated with fundamental weights.}
    \centering
   \begin{center}
\resizebox{9cm}{!}{\begin{tikzpicture}{xscale=1cm,yscale=1cm}
\coordinate[label=below:${\bf 133}$,label=above:$1$](B) at (-3.6,0);
\coordinate[label=below:${\bf 8645}$,label=above:$2$](C) at (-1.8,0);
\coordinate[label=below:${\bf 365750}$, label={[label distance=.6mm]45:$3$}](D) at (0,0);
\coordinate[label=below:${\bf 27664}$,label=above:$4$](E) at (1.8,0);
\coordinate[label=below:${\bf 1539}$,label=above:$5$](F) at (3.6,0);
\coordinate[label=below:${\bf 56}$,label=above:$6$](G) at (5.4,0);
\coordinate[label=left:${\bf 912}$,label=right:$7$](H) at (0,1.8);
\draw (B)--(C)--(D)--(E)--(F)--(G);
\draw (D)--(H);
\fill (B) circle (.1);
\fill (C) circle (.1);
\fill (D) circle (.1);
\fill (E) circle (.1);
\fill (F) circle (.1);
\fill (G) circle (.1);
\fill (H) circle (.1);
\end{tikzpicture}}
\end{center}
   
    \label{fig:E7dynkin}
\end{figure}
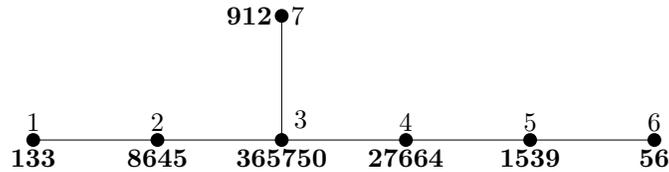
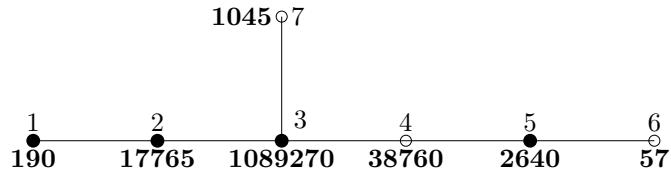
\begin{figure}[h]
 \caption{The analogy of Dynkin diagram for $\Eh$ and irreducible representations associated with fundamental weights. We call the three circled nodes \textit{fermionic fundamental weights}.}
    \centering
   \begin{center}
\resizebox{9cm}{!}{\begin{tikzpicture}{xscale=1cm,yscale=1cm}
\coordinate[label=below:${\bf 190}$,label=above:$1$](B) at (-3.6,0);
\coordinate[label=below:${\bf 17765}$,label=above:$2$](C) at (-1.8,0);
\coordinate[label=below:${\bf  1089270}$, label={[label distance=.6mm]45:$3$}](D) at (0,0);
\coordinate[label=below:${\bf  38760}$,label=above:$4$](E) at (1.8,0);
\coordinate[label=below:${\bf  2640}$,label=above:$5$](F) at (3.6,0);
\coordinate[label=below:${\bf  57}$,label=above:$6$](G) at (5.4,0);
\coordinate[label=left:${\bf 1045}$,label=right:$7$](H) at (0,1.8);
\draw (B)--(C)--(D)--(E)--(F)--(G);
\draw (D)--(H);
\fill (B) circle (.1);
\fill (C) circle (.1);
\fill (D) circle (.1);
\draw (E) circle (.08);
\fill (F) circle (.1);
\draw (G) circle (.08);
\draw (H) circle (.08);
\end{tikzpicture}}
\end{center}
   
    \label{fig:E75dynkin}
\end{figure}

\subsection{Quadratic Casimir invariants}
We introduce the set of fermionic fundamental weights as $S_{\rm odd}=\{w_4,w_6,w_7\}$. 
We conjecture the following universal formula for the quadratic Casimir invariants of irreducible representations of $\Eh$:
\begin{align}\label{eq:C2}
C_2(R_\lambda)=\langle\lambda+2\rho,\lambda\rangle_{E_7}+\langle\lambda,\lambda\rangle_{\rm odd},
\end{align}
where the bilinear form $\langle , \rangle_{\rm odd}$ is defined by the matrix
\begin{align}\label{eq:oddcorrection}
M_{ij}=\begin{cases}
    1/2, & i,j\in S_{\rm odd},\, i=j, \\
    -1/2, &  i,j\in S_{\rm odd},\,i\neq j,\\
    0, & \mathrm{otherwise}.
\end{cases}.
\end{align}
More explicitly, we have
\begin{align}\nonumber
  \phantom{--,} C_2(R_\lambda)=&\,2 [ n_1 (n_1+3 n_2 + 4 n_3 + 3 n_4 + 2 n_5 + n_6 + 2 n_7 + 23)+ 
   n_2 (3 n_2+8 n_3 + 6 n_4+ 4 n_5 \\ 
   &   + 2 n_6 + 4 n_7 + 45)+ n_3(6 n_3 + 9  n_4 + 
   6  n_5  + 3  n_6 + 6  n_7 + 66)  + n_4(4 n_4 + 5  n_5 \\ \nonumber
   &  + 2  n_6 + 
   4  n_7 + 52 )
    + n_5(2 n_5 + 2  n_6 + 3  n_7 + 36 ) + n_6(n_6 + 
    n_7 + 19) + 2 n_7^2 + 34 n_7 ]
. 
\end{align}
Clearly, the quadratic Casimir invariant $C_2$ of arbitrary irreducible representation of $\Eh$ is even. The conjectural formula \eqref{eq:C2} successfully reproduces all known $C_2$ invariants for level 1 and 2 irreducible representations discussed in Section \ref{sec:level2}. It also passes many checks for higher level irreducible representations to produce indices that are always a multiple of 6. This entangles with the study on the representation dimensions, which will be discussed in more detail in Section \ref{sec:weyl}. One more evidence is that the conjectural $C_2$ formula \eqref{eq:C2} along with \eqref{eq:ldef} results in exactly the same index $l(k)$  for $(\Eh)_k$ with the one for $(E_7)_k$ for $k=3,4,5$, which are summarized in Conjecture \ref{conj1}. We believe these are not coincidences. We collect the quadratic Casimir invariants for both $E_7$ and $\Eh$ irreducible representations relevant to the current paper in Tables \ref{tb:reps4}, \ref{tb:reps5} and \ref{tb:repsCdM6}.

Naively, one may hope to have a bilinear form $\langle , \rangle_{\Eh}$ such that the above $C_2$ \eqref{eq:C2} is just $\langle\lambda+2\rho,\lambda\rangle_{\Eh} $. 
Unfortunately, we could not find a proper definition of such a form.

\subsection{Rational affine VOA $\Eh$ at levels $3,4,5$}
Assuming affine VOA $\Eh$ at level $k$ is rational, and the conjectural conformal weight formula \eqref{eq:conjch} and $C_2$ formula \eqref{eq:C2} hold, we can easily compute all conformal weights. Affine VOA $\Eh$ at level 3 has central charge $c=\frac{190}{9}$. We predict that there exist 12 conformal primaries and the conformal weights from small to big are
\begin{align}\label{eq:weights3}
    h_i=\left\{0,\frac{20}{27},\frac{8}{9},\frac{4}{3},\frac{38}{27},\frac{14}{9},\frac{5}{3},\frac{16}{9},\frac{56}{27},\frac{19}{9},\frac{20}{9},\frac{22}{9}\right\}.
\end{align}
The conductor\footnote{The g.c.d of the denominators of exponents $\alpha_i$.} is $N=108$. We do not find a proper Hecke image interpretation for the characters of this VOA. A significant difference from affine VOA $(E_7)_3$ is that the latter has a non-diagonal modular invariant, while $(\Eh)_3$ apparently cannot have. 

Similar with the level 2 case, all 12 (or less) conformal weights of affine VOA $(\mathfrak{g})_3$ for $\mathfrak{g}$ belonging to the Deligne-Cvitanovi\'c exceptional series \eqref{DC} and $\Eh$ can be obtained from some simple rational functions of dual Coxeter number $h^\vee$.  We collect the results in \eqref{DC3cw}.

Affine VOA $\Eh$ at level 4 has central charge $c=\frac{190}{7}$. We predict that there exist 25 conformal primaries and the conformal weights from small to big are
\begin{align}\label{eq:weights4}
    h_i=\left\{0,\frac{5}{7},\frac{6}{7},\frac{9}{7},\frac{19}{14},\frac{3}{2},\frac{45}{28},\frac{12}{7},\frac{25}{14},2,\frac{57}{28},\frac{15}{7},\frac{31}{14},\frac{16}{7},\frac{33}{14},\frac{17}{7},\frac{5}{2},\frac{18}{7},\frac{19}{7},\frac{11}{4},\frac{39}{14},\frac{20}{7},\frac{20}{7},3,\frac{23}{7}\right\} .
\end{align}
The conductor is $N=84$. Due the appearance of weight-$3/2$ primary, we suspect this theory may be fermionizable and possibly can be realized as certain fermionic Hecke image of a $c=\frac{10}{7}$ fermionic RCFT \cite{Lee:2022yic}. We remark that affine VOA $(E_6)_4$ allows fermionization \cite{Bae:2020xzl} and can be realized by fermionic Hecke operator \cite[Section 4.5]{Lee:2022yic}. 

Affine VOA $\Eh$ at level 5 has central charge $c=\frac{950}{29}$. We predict that there exist 44 conformal primaries and the conformal weights from small to big are
\begin{align}\label{eq:weights5}
    h_i= \bigg\{&0,\frac{20}{29},\frac{24}{29},\frac{36}{29},\frac{38}{29},\frac{42}{29},\frac{45}{29},\frac{48}{29},\frac{50}{29},\frac{56}{29},\frac{57}{29},\frac{60}{29},\frac{62}{29},\frac{64}{29},\frac{66}{29},\frac{68}{29},\frac{70}{29},\frac{72}{29},\frac{72}{29},\frac{75}{29},\frac{76}{29},\frac{77}{29},\frac{78}{29},\\& \nonumber
   \frac{80}{29},\frac{80}{29},\frac{83}{29},\frac{84}{29},\frac{84}{29},3,\frac{88}{29},\frac{90}{29},\frac{92}{29},\frac{93}{29},\frac{94}{29},\frac{95}{29},\frac{96}{29},\frac{98}{29},\frac{99}{29},\frac{100}{29},\frac{102}{29},\frac{104}{29},\frac{105}{29},\frac{110}{29},\frac{120}{29} \bigg\}.
\end{align}
The conductor is $N=348$. We expect this theory can have some non-diagonal modular invariants.

\subsection{Irrationality of affine VOA $\Eh$ at level higher than $5$}\label{sec:irr}
If affine VOA $\Eh$ at level  $k>5$ exists and is rational, then all characters of $(\Eh)_k$ should become a vector-valued modular form and satisfy a MLDE with \textit{integer index $l$}. However, assuming our conjectural formula \eqref{eq:C2} on the quadratic Casimir invariants is correct, it is easy to compute all the exponents and then the index $l$. We collect the computational results in the following table.
\begin{table}[h]
\def\arraystretch{1.1}
  \centering
  \begin{tabular}{c|c|c|c|c|c}\hline
 &  $k$  & $6$ & $7$ & $8$ &   $9$  \\ \hline
& $r(k)$  &  $79$ & $128$ & $208$ & $318$ \\ \hline
$E_7$ & $l(k)$ &  $ 2384 $ &  $6804  $ & $ 19015$ & $46056$ \\ \hline
$E_{7+1/2}$ & $l(k)$ &  $\frac{23837}{10}  $ &  $ \frac{210906}{31} $ & $\frac{152105}{8} $ &  $\frac{506574}{11}$ \\ \hline
     \end{tabular}
\end{table}

Clearly, these contradict with the fundamental property of MLDE that the index $l$ should be integers. This is strong evidence that the affine VOA $\Eh$ at level  $k>5$ (if it exists) is no longer rational.

\section{A conjectural Weyl dimensional formula}\label{sec:weyl}
It is an important question to determine all irreducible representations of intermediate Lie algebra $\Eh$, in particular all their dimensions. To this end, we study the Weyl-type dimension formula for $\Eh$. Our trial Weyl dimension formulas are only successful in two special situations, with no fermionic fundamental weight in which case we have \eqref{weyldim}, and with  one single fermionic fundamental weight in which case we have \eqref{weyldimodd}. 
Our two dimension formulas pass many consistency checks including
\begin{enumerate}
    \item For arbitrary integrable $\lambda\neq0$, the dimension $|R_{\lambda}|$ for $\Eh$ should be an integer larger than the $|R_{\lambda}|$ for $E_7$.
    \item For arbitrary $\lambda=n\theta$, the $|R_{\lambda}|$ should reproduce the dimension proved by Landsberg and Manivel \cite{LM}.
    \item The representation dimensions predicted by Cohen and de Man \cite{Cohen} should be recovered.
    \item The index $I$ for arbitrary $R_{\lambda}$ should be a multiple of $6$. This is inspired from the fact that all $E_7$ irreducible representations have indices as a multiple of $6$.
    \item The index should remain the same in arbitrary representation decomposition $\Eh \subset E_8$ and $E_7 \subset \Eh$.
\end{enumerate}

We collect all information we obtain on the $\Eh$ irreducible representations  with level $k\le 4$ in Table \ref{tb:reps4},  with level $k=5$ in Table \ref{tb:reps5}, and with higher levels that appeared in \cite{Cohen} in Table \ref{tb:repsCdM6}. As comparison, we also list the information on the  correspondent $E_7$ irreducible representations, i.e., with the same highest weight. In these Tables, for all $\Eh$ representations named by Cohen and de Man in \cite{Cohen}, we have marked their names. Except that the $-Y_5^*$ interpretation of the fundamental representation $\bf 57$ seems to be new, which will be discussed in Section \ref{sec:fdim}.  

We introduce a new concept to distinguish two types of irreducible representations. 
We first define \textit{bosonic weights} of $\Eh$ as the those weights with \textit{even} number of fermionic fundamental weights, and \textit{fermionic weights} as those with \textit{odd} number. Then we define \textit{bosonic representations} of $\Eh$ as the irreducible representations with bosonic highest weights, and  \textit{fermionic representations} as those with odd highest weights. One can think this as the typical quantum property that two fermions can pair as a boson. This concept turns out to be very useful when we discuss the properties of tensor product decompositions and flavored affine characters of $\Eh$. We use F/B to denote fermionic/bosonic representations in Tables \ref{tb:reps4}, \ref{tb:reps5} and \ref{tb:repsCdM6}. 
Notice that all $\Eh$ representations appearing in \cite{Cohen} are bosonic, except for the three negative ones $-Y_4^*,-G^*$ and $-H^*$. This is actually not surprising as all representations in \cite{Cohen} come from the tensor products of adjoint representation $\bf 190$ which is bosonic itself. The general rules of tensor product decompositions will  be  discussed  in Section \ref{sec:tensor}.

\begin{table}[ht]
\def\arraystretch{1.1}
\caption{All irreducible representations of $E_7$ and $E_{7+1/2}$ with level $k\le 4$. At each level, we order the weights by the $C_2$ of $E_{7+1/2}$.}
	\centering
	\begin{tabular}{|c|c|c|c|c|c|c|c|c|c|c|c|c|c|}
		\hline
$k$&	$\lambda$	   &  $E_7$  & $C_2$  & $I/6$  & $E_{7+1/2}$  &  $C_2$  & $I/6$ & CdM \cite{Cohen} & F/B \\
		\hline
$1$  &   $0000010$ & $\bf 56$ & $ \frac{57}{2} $ & $1$ & $\bf 57$ & $40$  & $1$ & $-Y_5^* $ & F \\ \hline
$2$  &     $1000000$ & $\bf 133$ & $ 36 $ & $3$ & $\bf 190$ & $ 48 $ & $4$ & $ \mathfrak{g}$ & B \\
$2$  &  $0000001$ & $\bf 912$ & $ \frac{105}{2} $ & $ 30$ & $\bf 1045$ & $ 72 $  & $ 33$ & $-Y_4^* $  & F  \\
$2$  &  $0000100$ & $\bf 1539$  & $ 56 $ & $54 $ & $\bf 2640$ & $ 76 $&  $ 88$ & $Y_2^* $  & B \\
$2$  &  $0000020$ & $\bf 1463$ & $ 60 $& $ 55 $  & $\bf 1520$ & $ 84 $&    $ 56 $ & $Y_3^* $ & B \\ \hline
$3$  &  $1000010 $ & $\bf 6480$ & $ \frac{133}{2} $& $ 270 $   & $\bf 9728$ & $90$&   $ 384 $  &  $-$ & F  \\
$3$  &  $0100000$ & $\bf 8645$ & $ 72 $& $ 390 $  & $\bf 17765$ & $ 96 $& $ 748 $  & $X_2$ & B \\
$3$  &  $0001000$ & $\bf 27664$ & $ \frac{165}{2} $& $ 1430 $  & $\bf 38760$ & $112  $   &$ 1904 $ & $-G^* $  & F  \\
$3$  &  $0000011$ & $\bf 40755$ & $ 84 $ & $ 2145 $ & $\bf 87040$ & $  114$    & $ 4352 $ & $C^* $ & B \\
$3$  &  $0000110$ & $\bf 51072$ & $ \frac{177}{2} $ & $ 2832 $   & {$\bf 102410$} & $120$   & $ 5390 $ & $ - $  & F \\
$3$  &  $0000030$ & $\bf 24320$ & $ \frac{189}{2} $& $ 1440 $  & $\bf 25840$ & $ 132 $  & $ 1496 $ & $-H^*  $  & F  \\ \hline
$4$  &  $2000000$ & $\bf 7371$ & $  76$ & $ 351 $ & $\bf 15504$ & $  100$      &$ 680 $ & $Y_2$ & B \\
$4$  &  $1000001$ & $\bf 86184$ & $ \frac{185}{2} $ & $ 4995 $ & $\bf 150480$ & $124  $  & $ 8184 $& $ - $  & F \\
$4$  &  $1000100$ & $\bf 152152$ & $ 96 $ & $9152  $ & $\bf 392445$ & $ 128 $   & $ 22032 $ & $A$ & B \\
$4$  &  $1000020$ & $\bf 150822$ & $ 100 $ & $ 9450 $ & $\bf 237405$ & $  136$    & $14161$ & $D^*$ & B\\
$4$  &  $0100010$ & $\bf 362880$ & $ \frac{209}{2} $& $ 23760 $  & $\bf 812592 $ & $ 140 $    & $49896  $& $ - $  & F \\
$4$  &  $0010000$ & $\bf 365750$ & $ 108 $ & $ 24750 $ & $\bf 1089270$ & $ 144 $  & $ 68796 $ & $X_3$ & B\\
$4$  &  $0000002$ & $\bf 253935$ & $ 112 $ & $17820  $ & $\bf 347490$ & $152$ &   $23166  $ & $ I^* $ & B  \\
$4$  &  $0000101$ & $\bf 861840$ & $ \frac{229}{2} $ & $61830$ & $\bf 1896960 $ & $ 154 $  & $ 128128 $& $ - $ & F\\
$4$  &  $0001010$ & $\bf 980343$ & $ 116 $ & $ 71253 $ & $\bf 3023280$ &   $  156$& $ 206856 $ & $F^*$ & B \\
$4$  &  $0000021$ & $\bf 885248$ & $ \frac{237}{2} $ & $65728$ & $\bf -$ &  $  160$&   $ - $& $ - $  & F \\
$4$  &  $0000200$ & $\bf 617253$ & $ 120 $ & $ 46410 $ & $\bf 2078505$ & $160$ &   $ 145860 $ & $ J $ & B  \\
$4$  &  $0000120$ & $\bf 915705$ & $ 124 $ & $71145  $ & $\bf -$ & $ 168 $  & $ - $ & $ - $ & B\\
$4$  &  $0000040$ & $\bf 293930$ & $ 132 $ & $24310$ & $\bf -$  & $184$&   $ - $& $ - $& B\\
		\hline
		\end{tabular}
			\label{tb:reps4}
		\end{table}

\begin{table}[ht]
\def\arraystretch{1.1}
	\centering
 \caption{All irreducible representations of $E_7$ and $E_{7+1/2}$ with level $k=5$. We order the weights by the $C_2$ of $E_{7+1/2}$.}
	\begin{tabular}{|c|c|c|c|c|c|c|c|c|c|c|c|c|c|}
		\hline
	$\lambda$	   &  $E_7$  & $C_2$  & $E_{7+1/2}$&  $C_2$ &    $I/6$   & CdM \cite{Cohen} & F/B \\
		\hline
$ 2000010 $   &  $\bf 320112 $   &  $ \frac{217}{2} $ & $\bf 723520$ & $ 144$   & $ 45696$  & $-$ & F \\  
$
 1100000 $   &  $\bf 573440 $   &  $ 114 $ & $\bf 1770496$ & 
 $ 150$   & $ 116480$ & $C$  & B \\  
 
 $ 1001000 $   &  $\bf 2282280 $   &  $ \frac{249}{2} $  & $\bf 4961280$ &  $166$  & $361216$ & $-$& F\\    $
 1000011 $   &  $\bf 3424256 $   &  $ 126 $  & $\bf 11316305$   & $168$ & $833833$ & $E$ & B\\  $
 0100001 $   &  $\bf 3792096 $   &  $ \frac{265}{2} $ & $\bf 10342080$ &  $176$ &  $798336$ & $-$& F\\  $
 1000110 $   &  $\bf 4522000 $   &  $ \frac{261}{2} $  & $\bf 13911040$ & $174$ &  $1061632$ & $-$& F\\  $
 0100100 $   &  $\bf 7142499 $   &  $ 136 $ & $\bf 28139760$ & $180$ &  $2221560$ & $F$ & B\\  
  $ 1000030 $   &  $\bf 2273920 $   &  $ \frac{273}{2} $  &$-$ &$186$ &$-$  &$-$ & F\\  
 $ 0100020 $   &  $\bf 7482618 $   &  $ 140 $  &  $-$&  $ 188$ &$-$ & $-$& B\\  
 
  $ 0010010 $   &  $\bf 13069056 $   &  $ \frac{285}{2} $ & $\bf44233728$ & $190$ &  $3686144$ & $-$& F\\   
 $ 0001001 $   &  $\bf 11316305 $   &  $ 144 $  & $\bf 49244580$ &    $ 192$ & $ 4146912$   & $X_4$& B\\ 
 $ 0000012 $   &  $\bf 9480240 $   &  $ \frac{293}{2} $ &$-$  &  $196 $&$-$ &  $-$& F\\ 
 $ 0001100 $   &  $\bf 14910896 $   &  $ \frac{297}{2} $ & $\bf 44651520$ & $198$ &  $3877632$ & $-$& F\\  
 $ 0000111 $   &  $\bf 23969792 $   &  $ 150 $ & $-$ &  $ 200$& $-$ & $-$ & B \\  
 $
 0001020 $   &  $\bf 17926272 $   &  $ \frac{305}{2} $  & $-$ &  $ 204$& $-$& $-$ & F\\  
  $ 0000210 $   &  $\bf 14220360 $   &  $ \frac{313}{2} $ & $\bf 59991360$ & $208$ &   $5472896$ & $-$ & F \\  
  $ 0000031 $   &  $\bf 12609597 $   &  $ 156 $  & $-$ &  $ 210$&$-$ & $-$ & B\\ 
 $ 0000130 $   &  $\bf 11376288 $   &  $ \frac{325}{2} $  & $-$ &  $220 $& $-$ &  $-$ & F \\  
 $ 0000050 $   &  $\bf 2785552 $   &  $ \frac{345}{2} $ & $-$& $240$ &$-$ & $-$ & F \\
		\hline
	
		\end{tabular}
			\label{tb:reps5}
		\end{table}

\begin{table}[ht]
\def\arraystretch{1.1}
\caption{Irreducible representations of $E_7$ and $E_{7+1/2}$ from \cite{Cohen} with level $k \ge 6$. All representations here are bosonic.}
	\centering
	\begin{tabular}{|c|c|c|c|c|c|c|c|c|c|c|c|c|c|}
		\hline
$k$&	$\lambda$	   &  $E_7$  & $C_2$  & $E_{7+1/2}$  &  $C_2$ &  $I/6$   & CdM \cite{Cohen}  \\
		\hline
$6 $  &  $3000000$ & $\bf 238602$ & $ 120 $ & $\bf 749360 $ & $ 156 $  & $ 51272  $ & $  Y_3$ \\
$6 $  &  $2000100$ & $\bf 6619239$ & $ 140 $ & $\bf 26001690 $    & $  184$& $ 2098382 $ & $ D  $ \\
$6 $  &  $0200000$ & $\bf 13728792$ & $ 156 $ & $\bf 64489040$   & $204  $ & $5770072   $& $  H$ \\
$ 6$  &  $1010000$ & $\bf 24386670$ & $ 152 $ & $\bf  111532869$   & $  200$ & $  9783585 $& $  I $ \\
$7 $  &  $2100000$ & $\bf 19046664$ & $160  $ & $\bf 89109240$ & $  208$ &   $  8129264 $& $ G $ \\
$ 8$  &  $4000000$ & $\bf 5248750$ & $ 168 $ & $\bf 24732110 $ & $  216$ & $2343042  $  & $ Y_4  $ \\
		\hline
	
		\end{tabular}
			\label{tb:repsCdM6}
		\end{table}

\subsection{Irreducible representations of  purely bosonic weights}
For any $\Eh$ weight $\lambda=\sum_{i=1,2,3,5}n_iw_i$, $n_{i}\in \NN$ which we call \textit{purely bosonic weights}, we find the following Weyl dimension formula for the $\Eh$ irreducible representations
\begin{equation}\label{weyldim}
\dim(R_\lambda)=\frac{\prod_{\alpha\in\Delta_+}\langle\lambda+\rho,\alpha\rangle\prod_{\alpha\in\frac{1}{2}\mathbf{56}}\langle\lambda+\rho,\alpha\rangle}{\prod_{\alpha\in\Delta_+}\langle\rho,\alpha\rangle\prod_{\alpha\in\frac{1}{2}\mathbf{56}}\langle\rho,\alpha\rangle}.
\end{equation}
The $\langle,\rangle$ is the $E_7$ bilinear form. 
We check that this formula always produces integer dimensions for purely bosonic weights. The resulting dimensions from small to large  are 
\begin{align}\nonumber
  &  1, 190, 2640, 15504, 17765, 392445, 749360, 1089270, 1770496, 
2078505, 24732110, 26001690,\\ \nonumber
&28139760, 64489040, 89109240, 111532869, 
252065970, 605537790, 737502480, 1050163440...
\end{align}
Assuming the conjectural formula \eqref{weyldim} is correct, we collect the representation dimensions with purely bosonic weights in Tables \ref{tb:reps4}, \ref{tb:reps5} and \ref{tb:repsCdM6}. They have a perfect match with the prediction of \cite{Cohen}.

The Weyl dimension formula \eqref{weyldim} is in fact the analogy of the Theorem 3.2 of \cite{LM02} which applies to $D_4,F_4$ and $E_6,E_7,E_8$ there.    When $n_2=n_3=n_5=0$, \eqref{weyldim} reduces to the proved dimension formula for $n\theta$ representations \cite[Theorem 7.1]{LM}, where $\theta$ is the fundamental weight generating the adjoint representation.  To be precise, we find
\begin{equation}
   \dim(n\theta)=\frac{(2n+23)}{2^{39}
3^{20}
5^9
7^6
11^4
13^3
17^2
19^1
23^1}\prod_{j=1}^{22}(n+j)\prod_{j=5}^{18}(n+j)\prod_{j=8}^{15}(n+j) 
\end{equation}
It is easy to check this is equivalent to the $a=6$ case of \cite[Theorem 7.1]{LM}.

\subsection{Irreducible representations with a fermionic fundamental weight}\label{sec:fdim}
When fermionic fundamental weights are involved, the original Weyl dimension formula \eqref{weyldim} does not produce  integer values. 
In general, we did not find a proper modified formula for the representation dimension with arbitrary weight. Nevertheless, when there is only one single fermionic fundamental weight, i.e., $n_4+n_6+n_7=1$, $n_{1,2,3,5}\in \NN$, we do find a reasonable dimension formula that always gives positive integer dimensions and reproduces all relevant representation dimensions predicted by Cohen and de Man \cite{Cohen}. 
We conjecture that for any weight $\lambda=\sum_{i=1}^7n_iw_i$ with $n_4+n_6+n_7=1$, the representation dimension is 
\begin{equation}\label{weyldimodd}
\dim(R_\lambda)=\frac{\prod_{\alpha\in\Delta_+}\langle\lambda+\rho,\alpha\rangle\prod_{\alpha\in\frac{1}{2}\mathbf{56}}\big(\langle\lambda+\rho,\alpha\rangle-\langle\lambda,\alpha\rangle_{\rm odd})}{2\prod_{\alpha\in\Delta_+}\langle\rho,\alpha\rangle\prod_{\alpha\in\frac{1}{2}\mathbf{56}}\langle\rho,\alpha\rangle}.
\end{equation}
Here $\langle , \rangle_{\rm odd}$ is the correction bilinear form defined in \eqref{eq:oddcorrection}, which makes $\langle\lambda+\rho,\alpha\rangle-\langle\lambda,\alpha\rangle_{\rm odd}$ an integer for all $\alpha\in\frac{1}{2}\mathbf{56}$. Note there is an extra 2 in the denominator, which may be related to the fermionic nature of this type of representations. We find for $\lambda=w_6,w_7,w_4$, the above formula produces $57$, $1045$ and $38760$ respectively, which are the dimensions of fundamental representation and representations $-Y_4^*$ and $-G^*$ in \cite{Cohen}. We believe these are not coincidences. Assuming the conjectural formula \eqref{weyldimodd} is correct, we collect relevant representation dimensions in Tables \ref{tb:reps4} and \ref{tb:reps5}. The first two new representations we predict are $\bf 9728$ and $\bf 102410$, both are level-3 fermionic representations. For representations with more than one fermionic fundamental weight, such as $\bf 1520$, unfortunately we could not find a Weyl-type formula to reproduce their dimensions.

As a side remark, we make an interesting observation that the fundamental representation $R_{w_6}=\mathbf{57}$ can actually be regarded as a new $-Y_5^*$ representation following the name rules of Cohen-de Man \cite{Cohen}. The dimension of $Y_k=k\theta$ representations for the Deligne-Cvitanovi\'c exceptional series including $\Eh$ was given by Landsberg-Manivel in \cite{LM02} (up to a negative sign) as
\begin{align}
    \dim Y_k= -\frac{(2k-1)\lambda-6}{k!\lambda^k(\lambda+6)}\prod_{i=1}^k\frac{((i-1)\lambda-4)((i-2)\lambda-5)((i-2)\lambda-6)}{(i\lambda-1)((i-1)\lambda-2)},
\end{align}
where $\lambda=-6/h^\vee$. It was noticed in \cite{Cohen} that the map $\lambda^*=1-\lambda$ always induce another Deligne dimension formula. Denote the representation induced from $Y_k$ by such map as $Y_k^*$. 
Then for e.g. $\mathfrak{g}=F_4,E_6,E_7,\Eh,E_8$, we compute the $Y_5^*$ representation as $0,\mathbf{1},0,-\mathbf{57},-\mathbf{248}$. In the same spirit, the $Y_6^*$ representation are $0,0,0,0,-\mathbf{1}$. Even higher $Y_k^*$ representations always vanish. In comparison, $Y_3^*$ of $G_2$ is its fundamental representation $\mathbf{7}$ and $Y_4^*$ of $F_4$ is its fundamental representation $\mathbf{26}$.

As special cases of \eqref{weyldimodd}, for $n\theta+w_6$ representations, we find 
\begin{equation}
\dim (n\theta+w_6)=\frac{1}{2^{36}
3^{19}
5^9
7^6
11^4
13^3
17^2
19^1
23^1
}\prod_{j=1}^{23}(n+j)\prod_{j=5}^{19}(n+j)\prod_{j=9}^{15}(n+j).
\end{equation}
The first few dimensions are $57,9728,723520,32248320,990880020,22764833280...$ For $n\theta+w_7$ representations, we find 
\begin{equation}
\dim (n\theta+w_7)=\frac{2n+25}{2^{39}
3^{19}
5^9
7^6
11^3
13^3
17^2
19^1
23^1
}\prod_{j=1}^{24}(n+j)\prod_{j=6}^{19}(n+j)\prod_{j=10}^{15}(n+j) .
\end{equation}
The first few dimensions are $1045,150480,9741680,386601930,10759940730...$ For $n\theta+w_4$ representations, we find 
\begin{equation}
\dim (n\theta+w_4)=\frac{n+13}{2^{36}
3^{18}
5^9
7^5
11^4
13^3
17^2
19^1
23^1
} \prod_{j=1}^{25}(n+j)\prod_{j=7}^{19}(n+j)\prod_{j=9}^{11}(n+j)\prod_{j=15}^{17}(n+j) .
\end{equation}
The first few dimensions are $38760,4961280,290801745,10596418560...$

\subsection{Decomposition $E_7 \subset\Eh \subset E_8$}\label{sec:filtration}
First consider the decomposition $E_7\subset E_{7+1/2}$. We make the basic assumption that
\begin{align}\label{assumption1}
R_{\lambda}^{E_{7+1/2}}=R_{\lambda}^{E_{7}}+\sum_{C_2(\mu)<C_2(\lambda),\, l(\mu)\le l(\lambda)}m_{\mu} R_{\mu}^{E_{7}},\qquad 0\le m_\mu\le 2.
\end{align}
Here $C_2$ and $l$ are the quadratic Casimir invariant and level for $E_7$ representations. In fact, mostly $0\le m_\mu\le1$. This reflects the key feature of multiplicity-free filtration $\mathfrak{g}_{r-1}\subset\mathfrak{g}_{r-1/2}\subset\mathfrak{g}_{r}$ for the classical types studied by Shtepin \cite{ShtepinAC,ShtepinB,ShtepinD0,ShtepinD}. However, we notice that for some $\Eh$ representations, there will be no possible decomposition to $E_7$ if restricting $0\le m_\mu\le1$ for all $m_\mu$. Therefore, we release the restriction to  $0\le m_\mu\le2$, which turns out to have a unique solution for all $\Eh$ representations under consideration. We believe this assumption \eqref{assumption1} on the decomposition $E_7\subset E_{7+1/2}$ should be valid at least for small and low level $\Eh$ representations.  

The two sides of \eqref{assumption1} are bound to have the same dimensions and indices. By explicitly scanning all possibilities, 
we are able to \textit{uniquely} determine lots of decompositions under $E_7 \subset \Eh$: for level 1 and 2, we have presented them in \eqref{eq:level1decomp}, \eqref{eq:level1decomp2} and \eqref{eq:level2decomp}--\eqref{eq:level2decomp3}, where all multiplicities $m_\mu$ satisfy $0\le m_\mu\le1$.
For level $3$ representations, we find the following unique decomposition:
\begin{align}
\bf  9728 &= {\bf 6480+ 1539+1463+ 133}+ 2\cdot\mathbf{56}+\bf 1  ,\\
\bf  17765 &= \bf 8645 + 6480 +1539+ 912  + 133 + 56 ,\\
\bf  38760 &= \bf 27664+8645+1539+912 ,\\
\bf 87040  &= {\bf 40755+ 27664+ 8645+ 6480+1539}+2\cdot \mathbf{912}+\bf 133 ,\\
\bf  102410 &= \bf 51072+40755+6480+1539+1463+912+133+56 ,\\
\bf 25840  &= \bf 24320+1463+56+1 .
\end{align}
Notice that the multiplicity 2 begins to appear for $\bf 9728$ and $\bf 87040$, which are fermionic and bosonic representations respectively.  
For level $4$ representations, we have the following unique decomposition:
\begin{align}
\bf  15504 & = \bf 7371+6480+1463+133+56+1 \\
 \bf  150480  &= \bf   86184+ 7371+ 40755+ 8645+ 6480+ 912+ 133,\\
 \bf  392445  &= {\bf 152152+ 86184+ 7371+ 51072+ 40755+ 27664+ 8645}+ 2\cdot\bf 6480
\\&\nonumber\ \ \  + \mathbf{1463}+2\cdot\bf 1539+ 
912+ 133+ 56
 \\
  \bf  237405  &= {\bf 150822+ 24320+ 51072+ 6480}+2\cdot {\bf1463+ 1539+ 133}+2\cdot\bf{ 56+ 1},\\
       \bf  812592  &= \bf 362880+ 150822+ 152152+ 51072+ 40755+ 27664+ 8645\\\nonumber&\ \ \ + 2\cdot {\bf6480+ 1463}+ 2\cdot \bf{1539+
912+ 133+ 56}
,\\
        \bf  1089270  &= {\bf 365750+ 362880+ 152152+ 86184+ 40755}+ 2\cdot{\bf 27664}+ 2\cdot \bf 8645\\\nonumber&\ \ \ \bf   + 6480+ 1539+ 912,\\
         \bf   347490 &= \bf 253935+ 86184+ 7371,\\
         \bf   1896960 &= {\bf 861840+ 253935+ 365750+ 152152}+ 2\cdot\bf 86184+ 7371+ 40755\\ \nonumber
         &\ \ \ \bf+ 27664+ 8645+ 6480
,\\
\bf   3023280 &= \bf 980343+ 861840+ 365750+ 362880+ 152152+ 86184+ 51072\\ \nonumber
           &\ \ \ + 2\cdot\mathbf{ 40755}+ 2\cdot \mathbf{27664}+ \
2\cdot \mathbf{8645}+\bf 6480+ 1539+ 912.
\end{align}
There are three remaining irreducible representations marked in Table \ref{tb:reps4} which we could not determine the decompositions as we do not know their dimensions. 
For level 5, we can only determine the decomposition of two irreducible representations 
\begin{align}
            \bf   723520 &= {\bf 320112+ 150822+ 152152+ 7371+ 24320+ 51072}+ 2\cdot\mathbf{6480}\\ \nonumber
            &\ \ \ + 2\cdot{\bf 1463+ 1539+ 133}+ 2\cdot\bf 56+1,\\
\bf   1770496 &= \bf 573440+ 362880+ 320112+ 150822+ 152152+ 86184+ 51072\\\nonumber 
&\ \ \ +{\bf  40755+ 
8645+7371}+ 2\cdot \bf 6480+ 1463+ 1539+ 912+ 133+ 56.
\end{align}
For level 6, we can only determine the decomposition of one smallest irreducible representation 
\begin{align}
           \bf   749360 &= \bf 320112+ 238602+ 150822+ 24320+ 7371\\ \nonumber
           &\ \ \ \bf+ 6480+ 1463+ 133+ 56+ 1.
\end{align}
It would be desirable to understand the factor 2 appearing here and there in the above decompositions.

For $\Eh \subset E_8$ decomposition, our strategy is to first look at the decomposition $E_7\times SU(2)\subset E_8$ and  find the $E_7$ representation $R_\lambda$ with the largest $C_2$. Then it is reasonable to assume that in $E_{7+1/2}\subset E_8$ decomposition, the highest $E_{7+1/2}$ representation also has  weight $\lambda$.
Once we know the $ E_7\times SU(2)\subset E_8$ and all the relevant $E_7\subset E_{7+1/2}$ decompositions, we can do the deduction from high $C_2$ to low $C_2$ one by one to find the precise $E_{7+1/2}\subset E_8$ decomposition. Indeed, we are able to uniquely determine the following decompositions under $\Eh \subset E_8$ for the first 11 irreducible representations of $E_8$ as follows.
\begin{align}
\bf 248&=\bf 190+57+1,\\
\bf3875 &=\bf2640 + 1045+190 ,\\
\bf 27000&=\bf 15504+9728 + 1520 + 190 + 57 + 1,\\
 \bf   30380& =\bf 17765+ 9728+ 2640+  190+ 57,\\
 \bf 147250 &=\bf 87040+ 38760+ 17765+ 2640+ 1045,\\
 \bf 779247&=\bf 392445+ 150480+ 102410+ 87040+ 17765+ 15504
\\ \nonumber
 &\phantom{=}\bf\ + 9728+ 2640+ 1045+ 190,\\
 \bf 1763125  &=\bf 749360+723520+237405+25840+15504+9728\\ \nonumber&\ \ \ \bf+1520+190+57+1,\\
  \bf 2450240&=\bf 1089270+812592+392445+87040+38760+17765\\ \nonumber&\  \  \ \bf+9728+2640,\\
  \bf 4096000&=\bf 1770496 + 723520 + 812592 + 392445 + 237405 + 102410  \\ \nonumber
  &\ \ \ + \mathbf{15504+ 17765} 
+{ 2}\cdot \bf 9728 + 2640 + 1520 + 190 + 57
,\\
\bf 4881384&=\bf 2078505 + 1896960 + 347490 + 392445 + 150480 + 15504,\\
 \bf 6696000&=\bf 3023280 + 1896960 + 1089270 + 392445 + 150480 + 87040\\ \nonumber
 &\ \ \ \bf  + 38760 + 17765.
\end{align}
For each decomposition, we have checked that the two sides have the same indices. We also checked the consistency with both sides decomposed to $E_7$ irreducible representations. It is interesting to remark that the multiplicity is not always one, e.g. for $\bf 4096000$, which is the only level-$5$ irreducible representation of $E_8$ among the above 11 ones. 

The next $E_8$ irreducible representation $\bf 26411008$ should involve several $E_{7+1/2}$ representations that we do not know the dimension or decompositions to $E_7$, thus it is not possible to determine its $\Eh \subset E_8$ decomposition based on the current data.

\subsection{Tensor product decompositions}\label{sec:tensor}
Unlike the tensor product decomposition for simple Lie algebras, we notice that for intermediate Lie algebra $\Eh$, the tensor product decomposition of irreducible representations often involves negative signatures. One reason for this unorthodox phenomenon might be that there actually exist  more representations of $\Eh$ than we anticipate. However, we will not pursue this possibility in the current paper. 

Some tensor product decompositions involving only  bosonic representations have been studied in \cite{Cohen}, in particular the tensor products of (up to four) adjoint representations. For example, for  $\bf 190$ of $\Eh$, \cite{Cohen} gave
\begin{align}
   \textrm{Sym}^2 \bf  190 &=\bf 15504  + 2640  + 1,\\
  \textrm{Alt}^2\bf 190 &=\bf  17765 + 2640 .
\end{align}
We will focus on the tensor products involving fermionic representations. 

To compute the tensor product decomposition of $\Eh$ representations $R_\lambda$ and $R_\mu$, we first convert $R_\lambda$ and $R_\mu$ into $E_7$ irreducible representations under $E_7\subset \Eh$, then compute the tensor product decompositions of all pairs of $E_7$ irreducible representations, and finally convert the result back to $\Eh$ irreducible representations following the same deduction method used in Section \ref{sec:filtration} for $\Eh\subset E_8$ decompositions. Clearly this process is unique.  
For example, by this method we are able to determine the  tensor product decompositions of fundamental representation $\bf 57$ as
\begin{align}
\textrm{Sym}^2\bf 57&= \bf  1520+190-57  ,\\
\textrm{Alt}^2\bf 57&= \bf  2640-1045+1  .
\end{align}
For the fermionic representation $\bf 1045$, we find its tensor product decompositions as
\begin{align}
\textrm{Sym}^2\bf 1045&= \bf347490+392445-150480-102410+87040-38760\\ \nonumber
&\phantom{=}\ \mathbf{ +17765-9728}+2\cdot \mathbf{1520+190-57},\\
\textrm{Alt}^2\bf 1045&= \bf 1089270 - 812592 + 237405+ 15504 - 25840+87040\\ \nonumber
&\phantom{=}\ \mathbf{-38760-9728}+2\cdot \mathbf{2640}-2\cdot \mathbf{1045+1}.
\end{align}

Moreover, 
we obtain the following tensor product decompositions involving the fundamental representation $\bf 57$:
\begin{align}
  \bf 190\times 57&=\bf    9728+ 1045+57,\\  
\bf 1045 \times 57&=\bf      87040 + 17765+ 2640+ 1520+190 - 38760 - 9728  - 1045 - 57 ,\\
\bf 2640 \times 57&=\bf  102410 + 38760 + 9728 + 1045 + 57-1520 ,\\
\bf 1520 \times 57&=\bf 102410 + 38760+ 25840 + 9728 + 1045+ 57 - 87040  - 2640  - 1520  ,\\
\bf 9728 \times 57&=\bf 392445+237405    + 87040  + 17765+ 15504  + 
 2640 +1520 + 190\\ \nonumber
 &\ \ \  \bf - 150480- 38760 - 9728 - 1045   ,\\
\bf 17765 \times 57&=\bf 812592 + 150480 + 38760 + 9728 + 1045,\\
\bf 38760 \times 57&=\bf 3023280  - 1896960 + 347490 + 1089270 - 812592 + 237405\\  \nonumber
&\ \ \  +{\bf  392445 - 
150480 - 25840 - 102410} + 2\cdot \mathbf{87040} - 
 2\cdot \mathbf{38760}\\ \nonumber
 &\ \ \ \bf+ 17765 - 9728 + 2640 + 1520  .
\end{align}
We also find the following decompositions of tensor products between an irreducible representation and the adjoint representation
\begin{align}
  \bf 1045  \times 190 &=\bf 150480 + 38760 + 9728 + 1045 + 57 - 1520,\\
\bf 1520\times 190 & =\bf 237405 + 87040  + 2640  + 1520 - 38760 - 1045.
\end{align}

Recall we use F/B to denote a fermionic/bosonic irreducible representation of $\Eh$. 
Based on the above explicit computations, we observe that the tensor product decomposition of type $B\times B$ or $F\times F$ has all bosonic representations with positive signs and all fermionic ones with negative signs. On the other hand,  the tensor product decomposition of type $B\times F$ has all fermionic representations with positive signs and all bosonic ones with negative signs. These resemble the coupling between bosons and fermions in quantum mechanics except for the negative part.

\subsection{Flavored affine characters}
With the decomposition data of $\Eh$ irreducible representations in Section \ref{sec:filtration}, formally we are able to determine the following decomposition of the Fourier coefficients of the flavored characters of VOA $(\Eh)_1$. For the vacuum character, we find
\begin{align}\label{eq:chilevel1a}
    \chi_0=\,&q^{-\frac{19}{60}}(1+\mathbf{190} q+(\mathbf{1 + 190 + 2640}) q^2+(\mathbf{1 + 17765 + 2640 + 1520} + 2\cdot \mathbf{190}) q^3\\ \nonumber
& +(2+\mathbf{17765 + 87040} + 3\cdot\mathbf{ 2640 + 1520 + 15504} + 3\cdot\mathbf{ 190 - 1045}) q^4\\ \nonumber
&+(\mathbf{392445 + 15504} + 2\cdot \mathbf{87040}  + 3\cdot \mathbf{17765} + 3\cdot \mathbf{1520} + 4\cdot \mathbf{2640} + 
 6\cdot \mathbf{190} + 2\\\nonumber
 &  -\mathbf{ 38760- 1045  - 57})q^5+(\mathbf{1089270 + 237405} + 2\cdot \mathbf{392445} + 4\cdot \mathbf{87040} \\ \nonumber & + 3\cdot \mathbf{15504} + 5\cdot \mathbf{17765} + 5\cdot \mathbf{1520} - 
 2\cdot \mathbf{38760} + 9 \cdot\mathbf{2640} + 8\cdot \mathbf{190} + 5 - \mathbf{25840}\\ \nonumber
 &-\mathbf{ 9728} - 3 \cdot\mathbf{1045 - 57})q^6+(\mathbf{1770496 + 3023280 + 1089270 + 347490}\\ \nonumber
 &+ 2\cdot \mathbf{237405} + 5\cdot \mathbf{392445} + 
  4\cdot \mathbf{15504}+ 9\cdot \mathbf{87040} + 10\cdot \mathbf{17765} + 10\cdot \mathbf{1520} \\ \nonumber
 &+ 13\cdot \mathbf{2640} + 15\cdot \mathbf{190} + 5 - 
  \mathbf{150480 - 25840- 102410} - 5 \cdot\mathbf{38760} - 2\cdot \mathbf{9728}\\ \nonumber
 & - 5\cdot \mathbf{1045} - 3\cdot \mathbf{57})q^7+\dots).
\end{align}
We have checked the above Fourier coefficients with further decomposition to $E_7$ irreducible representations. We notice that unlike the simple Lie algebra cases, the decomposition of the Fourier coefficients of $\Eh$ affine characters could contain negative signs. This may be not so surprising knowing that the affine $A_{1/2}$ characters also involve negative signs, which can be viewed as the supercharacters of affine Lie superalgebra $B_{0,1}$. 
Similarly for the non-vacuum character, we find
\begin{align}\label{eq:chilevel1b}
\chi_{4/5}=\,&q^{\frac{29}{60}}(\mathbf{57}+ (\mathbf{1045 + 57}) q + (\mathbf{1045 + 9728} + 2\cdot\mathbf{ 57 - 1520}) q^2 + (\mathbf{38760} + 2\cdot \mathbf{ 9728}\\ \nonumber
&\ + 3\cdot \mathbf{ 1045} + 3\cdot \mathbf{ 57} -  \mathbf{2640 - 1520}) q^3 + (\mathbf{150480 + 102410} + 2\cdot \mathbf{38760}\\ \nonumber
&\  + 4 \cdot\mathbf{9728}+ 5\cdot \mathbf{1045} + 6\cdot\mathbf{ 57} - \mathbf{87040} - 
 3\cdot \mathbf{1520 - 2640 - 190})q^4+(\mathbf{812592}\\ \nonumber
 &\ + 2\cdot \mathbf{150480 + 25840}+ 2\cdot \mathbf{102410} + 5\cdot \mathbf{38760} + 8\cdot \mathbf{9728} + 10\cdot \mathbf{1045} + 
 9\cdot \mathbf{57}\\ \nonumber
 &\ - \mathbf{237405} - 2\cdot \mathbf{87040} - \mathbf{17765} - 5\cdot \mathbf{1520 }- 3\cdot \mathbf{2640 - 190 - 1})q^5 +(\mathbf{723520}\\ \nonumber
 &\ + \mathbf{1896960} + 2\cdot \mathbf{812592} + 5\cdot \mathbf{150480} + 2\cdot \mathbf{25840} + 5\cdot \mathbf{102410} + 
 11\cdot \mathbf{38760} \\ \nonumber
 &\ + 15\cdot \mathbf{9728} + 16\cdot \mathbf{1045} + 16\cdot \mathbf{57} - \mathbf{347490} - 2 \cdot\mathbf{237405 - 392445} - 
 5 \cdot\mathbf{87040}\\ \nonumber
 &\ - 2\cdot \mathbf{17765} - 11\cdot \mathbf{1520} - 5\cdot \mathbf{2640} - 3\cdot \mathbf{190} - 1) q^6+\dots) .
\end{align}
Interestingly, we observe that for the vacuum character $\chi_0$, all $\Eh$ representations with positive signs are bosonic, while all those with negative signs are fermionic. By contrast, for the non-vacuum character $\chi_{4/5}$, all $\Eh$ representations with positive signs are fermionic, while all those with negative signs are bosonic. 

For the flavored characters of affine VOA $(\Eh)_2$, we find the following decompositions of their Fourier coefficients
\begin{align}
    \chi_0=\,&q^{-\frac{95}{156}} (\mathbf{1} + \mathbf{190} q + ({\bf 1+190+2640+ 15504}) q^2  +({\bf 1} + 3\cdot{\bf 190 + 1520 }\\ \nonumber
&\qquad\ \, +{\bf 2640+15504} + 
 2\cdot{\bf 17765 + 392445})q^3+\dots),\\
\chi_{\frac{10}{13}}=\,&q^{\frac{25}{156}} (\mathbf{57} + ({\bf 57 + 1045 + 9728}) q +(3\cdot\mathbf{ 57} + 2\cdot\mathbf{ 1045} + 3\cdot\mathbf{ 9728} + \mathbf{38760} \\\nonumber
&\qquad  +\mathbf{ 102410+ 150480 - 1520})q^2+\dots),\\
\chi_{\frac{12}{13}}=\,&q^{\frac{49}{156}} (\mathbf{190} + ({\bf 1+190+2640+ 17765}) q +({\bf 1} + 3\cdot\mathbf{190} + \mathbf{1520} + 2\cdot \mathbf{2640}\\\nonumber
&\qquad  + \mathbf{15504} + 
 2\cdot \mathbf{17765} + \mathbf{87040} + \mathbf{392445})q^2+\dots),\\
\chi_{\frac{18}{13}}=\,&q^{\frac{121}{156}} (\mathbf{1045} + ({\bf 57 + 1045 + 9728 + 38760 - 1520}) q +\dots),\\
\chi_{\frac{19}{13}}=\,&q^{\frac{133}{156}} (\mathbf{2640} + (\mathbf{87040 + 17765 + 1520 + 2640 + 190}) q +\dots),\\
\chi_{\frac{21}{13}}=\,&q^{\frac{157}{156}} (\mathbf{1520} + (\mathbf{1520 + 2640 + 87040 - 38760 - 1045}) q +\dots).
\end{align}
Similar to the level one case, it is easy to see that the affine characters $\chi_{10/13}$ and $\chi_{18/13}$ have fermionic nature, while  the rest four have bosonic nature. We conjecture that all $\Eh$ representations appearing in $\chi_{10/13}$ and $\chi_{18/13}$ with positive signs are fermionic, while those with negative signs are bosonic. The rest four characters have exactly the opposite property. Since these flavored characters involve negative Fourier coefficients, they might be viewed as supercharacters of some superalgebras related to $\Eh$.

\section{Rank-$n$ instanton VOA}
In this section, we turn to an entirely different type of VOAs called instanton VOAs, originating from the conjectural 4d SCFT/VOA correspondence \cite{Beem:2013sza}. In Section \ref{sec:rank1}, we have reviewed the rank-one instanton VOA $({\mathfrak{g}})_{-{h^\vee}/{6}-1}$ for the Deligne-Cvitanovi\'c exceptional series and $\Eh$. The higher-rank generalization, especially the rank-two instanton VOAs, for the Deligne-Cvitanovi\'c exceptional series have been proposed in \cite{Beem:2019snk}. A new feature for the higher-rank generalization is the appearance of an extra $SU(2)$ factor. The level for rank-$n$ instanton VOA for $\mathfrak{g}=A_1,A_2,D_4,E_6,E_7,E_8$ was given in \cite{Beem:2019snk} as 
\begin{align}
(\mathfrak{g})_{-n(h^\vee+6)/6}\times SU(2)_{-(n-1)(6 + n( h^\vee+6))/12}.
\end{align}
The central charge was given as
\begin{align}
c=-(6n^2+n(n+3)h^\vee-2)/2.
\end{align}

The explicit VOA constructions for the rank-two cases have been studied in \cite{Beem:2019snk}.
Remarkably, it was found that the vacuum characters of rank-two instanton VOAs for the Deligne-Cvitanovi\'c exceptional series satisfy an uniform fourth order \textit{twisted} modular linear differential equations with all numerical coefficients as rational functions of $h^\vee$. The {twisted} means that the coefficient functions of the MLDE are $\Gamma(2)$ modular forms, instead of $SL(2,\mathbb{Z})$ ones as in the positive level cases and the rank-one case. Suppose a conjectural rank-two VOA associated with $\Eh$ exists, we expect its vacuum character should satisfy a similar MLDE. 
The uniform fourth order twisted MLDE \cite[Equation (5.5)]{Beem:2019snk} by taking $h^\vee=24$ gives
\begin{align}
&\bigg[D^4  -\frac{11}{6}\Theta_{0,1} D^3  - \left(\frac{4705}{288}\Theta_{0,2}-\frac{2519}{288}\Theta_{1,1}\right)D^2
-\left(\frac{965}{1152}\Theta_{0,3}-\frac{28601}{1152}\Theta_{1,2}\right)D \\ \nonumber
&\ +\frac{4594825 \Theta_{0,4}}{331776}+\frac{18025993 \Theta_{1,3}}{82944}-\frac{25351319 \Theta_{2,2}}{110592}\bigg]\chi=0,
\end{align}
where
$\Theta_{r,s}(\tau)= \theta_2(\tau)^{4r}\theta_3(\tau)^{4s} + \theta_2(\tau)^{4s}\theta_3(\tau)^{4r}$.
From this MLDE, we solve the vacuum character of the conjectural rank-two VOA associated with $\Eh$ as 
\begin{align}
\chi_{vac}=&\,q^{131/24}(1+193 q+380 q^{3/2}+18914 q^2+68060 q^{5/2}+1299299 q^3+6168280 q^{7/2}\\ \nonumber
&+70763062 q^4+379716500 q^{9/2}+\dots).
\end{align}
We also find the flavored vacuum character up to the overall factor $q^{131/24}$ as
\begin{align}\label{eq:rank2f}
&1+(\mathbf{190}+\chi_3) q+\mathbf{190}\chi_2 q^{3/2}+(\chi_5 +(\mathbf{190} + 1)\chi_3+ \mathbf{15504}+ \mathbf{2640} + \mathbf{190} + 2) q^2\\ \nonumber
&+(\mathbf{190}\chi_4+(2\cdot\mathbf{15504}+ \mathbf{2640}+2)\chi_2)q^{5/2}+(\chi_7+(\mathbf{190}+1)\chi_5
+(2\cdot\mathbf{15504}+ \mathbf{2640}\\ \nonumber
&+2\cdot\mathbf{190}+4)\chi_3+\mathbf{749360+392445+15504+2640}+2\cdot\mathbf{17765}+4\cdot\mathbf{190}+2)q^3\\ \nonumber
&+(\mathbf{190}\chi_6+(\mathbf{17765}+\mathbf{15504}+3\cdot\mathbf{190}+1)\chi_4+(\mathbf{1770496+749360+392445}\\ \nonumber
&+2\cdot\mathbf{17765}+4\cdot\mathbf{15504}+2\cdot\mathbf{2640}+4\cdot\mathbf{190}+3)\chi_2)q^{7/2}+\dots.
\end{align}
Here $\chi_n$ is the character of the $n$-dimensional irreducible representation of $SU(2)$. This follows from a general ansatz for the Schur indices of rank-two instanton  SCFTs $H_{\mathfrak{g}}$ in \cite[Equation (5.26)]{Gu:2019dan}\footnote{In the notations of \cite[Equation (5.26) and Table 5]{Gu:2019dan}, the three relevant representations for $\Eh$ are $C_6=\bf 1520$, $C_7= \bf 392445$ and $B_2=\bf 1770496$.} or \cite[Equation (5.4)]{Beem:2019snk}. Notice that all irreducible representations of $\Eh$ appearing in \eqref{eq:rank2f} are bosonic. We regard the nice behavior of the above vacuum character as a strong support that rank-two VOA associated with $\Eh$  indeed exists.

Consider the VOAs associated with higher rank moduli space of instantons. From the general results for instanton VOA associated to Deligne-Cvitanovi\'c exceptional series \cite{Beem:2019snk}, it can be expected that the rank-$n$ instanton VOA associated with $\Eh$ has global symmetry
\begin{align}
    (\Eh)_{-5n}\times SU(2)_{-(5n+1)(n-1)/2}.
\end{align}
It would be interesting to determine the vacuum character of these higher-rank instanton VOAs. For example, it was conjectured in \cite{Beem:2019snk} that the Schur indices of rank-3 $H_{\mathfrak{g}}$ theories, conjecturally equivalent to the vacuum characters of rank-3 instanton VOAs, satisfy a uniform 7th order twisted MLDE. In particular, the 7th order MLDE for $\mathfrak{g}=D_4$ was given in \cite[Equation (6.2)]{Beem:2019snk}. Following a general ansatz for the Schur indices of rank-3 $H_{\mathfrak{g}}$ theories \cite[Equation (5.63)]{Gu:2019dan}, we expect the vacuum character of rank-3 instanton VOA associated to $\Eh$ has the following Fourier expansion (up to the overall factor) 
\begin{align}
\chi_{vac}=1 + 193q + 384 q^{3/2} +19485q^2   + 74496 q^{5/2} + 1430318q^3+\dots .
\end{align}
The flavored vacuum character is expected to be
\begin{align}\label{eq:rank3f} 
&1 + (\chi_3+\mathbf{190})q + (\chi_4+\mathbf{190}\chi_2) q^{3/2}  +(\chi_{5}+(2\cdot\mathbf{190}+1)\chi_3+\mathbf{15504+2640}+\mathbf{190}+3)q^2\\ \nonumber
&+ (\chi_6+2(\mathbf{190}+1)\chi_4+(2(\mathbf{15504+2640})+\mathbf{190}+3)\chi_2)q^{5/2}+ (2\chi_7+(3\cdot\mathbf{190}+1)\chi_5\\ \nonumber
&
 +(4\cdot\mathbf{15504}+3\cdot\mathbf{2640}+3\cdot \mathbf{190}+8)\chi_3+ \mathbf{749360+392445+17765+1520}\\ \nonumber
&+3(\mathbf{15504+2640})+2\cdot \mathbf{190}+5)q^3 +\dots .
\end{align}
Notice again that all irreducible representations of $\Eh$ appearing in \eqref{eq:rank3f} are bosonic.

It is intriguing to consider whether one can genuinely construct 4d $\mathcal{N}=2$ SCFTs in a certain sense corresponding to the rank-$n$ instanton VOAs associated with $\Eh$. They might even allow class S constructions from $(2,0)$ type $A_{5n-1}$ SCFTs, following the fact that 4d rank-$n$ instanton SCFTs of $D_4,E_6,E_7,E_8$ can be obtained from class S constructions of type $A_{2n-1}$, $A_{3n-1}$, $A_{4n-1}$, $A_{6n-1}$ respectively \cite{Benini:2009gi}. If class S constructions can be found, it would be interesting to compute the Schur indices, Hall-Littlewood indices and Macdonald indices of the SCFTs following \cite{Gadde:2011uv,Gaiotto:2012uq}.

\section{$\Eh$ as a gauge algebra}
In the rank-$k$ instanton SCFT/VOA correspondence, we have seen the possibility of $\Eh$ as a \textit{flavor algebra}. 
An even more intriguing question is whether it is possible to realize $\Eh$ as a \textit{gauge algebra}. We find that 
if $\Eh$ can be realized as a certain gauge algebra, then following a celebrated 3d monopole formula of Benvenuti-Hanany-Mekareeya \cite{Benvenuti:2010pq}, the \textit{5d one $\Eh$ instanton Nekrasov partition function}, i.e., the \textit{K-theoretic one $\Eh$ instanton Hilbert series} should be 
\begin{align}
Z_1^{\rm Nek}&=v^{h^\vee-1}\sum_{n=0}^\infty v^{2n}\chi_{n\theta}^{\mathfrak{g}}\\ \nonumber
&=v^{23}(1+\mathbf{190}v^2+\mathbf{15504}v^4+\mathbf{749360}v^6+\mathbf{24732110}v^8+\mathcal{O}(v^{10})).
\end{align}
Here we have factored out the center of motion contribution, such that the one-instanton partition function only depends on $v=e^{\epsilon_+}$ but not on $\epsilon_-$.  
If turning off the $\Eh$ gauge fugacities, the character $\chi_{n\theta}$ becomes the $\dim{n\theta}$ and we obtain the following rational expression for the infinite summation:
\begin{align}
    \frac{1}{(v-v^{-1})^{46}}(&v^{\pm23}+144 v^{\pm21}+7799 v^{\pm19}+217646 v^{\pm17}+3587175 v^{\pm15}+37732006 v^{\pm 13}\\[-1mm] \nonumber
&+266204829 v^{\pm11}+1303208244 v^{\pm9}+4533843651 v^{\pm7}+11399199625 v^{\pm5}\\ \nonumber
&+20952141111 v^{\pm3}+28356500429 v^{\pm1}),
\end{align}
where $v^{\pm n}$ is a short notation for $ v^n+v^{-n}$. Thus, 
this expression is palindromic with respect to $v$ as required by the symmetry of $Z_1^{\rm Nek}$. We regard this as strong evidence that a \textit{5d $\mathcal{N}=1$ $\Eh$ gauge theory} should exist in a certain sense. It is intriguing to consider whether there exists a compact formula for $Z_1^{\rm Nek}$ with $\Eh$  fugacities turned on. One possible approach is to use the correspondence between the K-theoretic Nekrasov partition function and Hall-Littlewood index, see e.g. \cite{Gadde:2011uv,Gaiotto:2012uq}.  Another possible approach is to generalize the derivation on the Hilbert series of the one-instanton moduli space for simple Lie algebras in \cite{Keller:2011ek}.

A 5d gauge theory can carry hypermultiplets in various representations. Carrying an adjoint hypermultiplet makes it a 5d $\mathcal{N}=1^*$ $\Eh$ gauge theory, which might allow a 6d description. We also expect a 5d $\mathcal{N}=1$ $\Eh$ gauge theory can carry at most four fundamental $\bf 57$ matters. It would be desirable to find the exact $v$ expansion formulas for the K-theoretic one instanton partition functions for these \textit{intermediate gauge theories}. Many such formulas for exceptional gauge theories with matters have been found in \cite{DelZotto:2018tcj,Gu:2020fem}.

One can even imagine $\Eh$ as a 6d gauge algebra. 
In particular, if a 6d $(1,0)$ theory exhibits pure $\Eh$ gauge symmetry, it should be geometrically engineered by a non-compact Calabi-Yau threefolds as certain elliptic fibration over $O_{\mathbb{P}^1}(-10)$.\footnote{This putative theory should not be confused with the 6d $(1,0)$ $E_7$ gauge theory with half-hypermultiplet $\frac12\mathbf{56}$ \cite{Morrison:2012np}, which is a well-known non-Higgsable cluster SCFT geometrically engineered by a non-compact Calabi-Yau threefolds as an  elliptic fibration over $O_{\mathbb{P}^1}(-7)$. The Mori cone of this Calabi-Yau geometry has been discussed in \cite{Haghighat:2014vxa}.} It seems that such base geometry could be realized by blowup down of a $(-12)$-curve intersecting with two $(-1)$-curves which engineers a 6d $(1,0)$ pure $E_8$ theory coupled with two E-string theories. The Mori cone of the Calabi-Yau geometry for the quiver theory has been discussed in \cite[Section 4.2.7]{Haghighat:2014vxa}. Besides, from \cite[Section 7.4.3]{Gu:2020fem}, we know such 6d quiver theory does not have unity blowup equations. Thus the elliptic genera of the self-dual strings are not easy to compute. One can further expect on $(-10+2n_f)$-curve, there may exist a 6d $(1,0)$ $\Eh$ gauge theory with $n_f$ number of fundamental $\bf 57$ hypermultiplets for $n_f=0,1,2,3,4$.

Furthermore, following a surprising conjecture of del Zotto and Lockhart on the relation between Schur index and elliptic genus \cite{DelZotto:2016pvm}, we expect the one-string elliptic genus of 6d $(1,0)$ pure $\Eh$ theory (if it exists) should have an exact relation with the vacuum character of VOA $(\Eh)_{-5}$ in \eqref{eq:level-5}. 

\section{The theta block associated with $\Eh$} 
Though we did not discuss the possible generalization of Weyl character formula or Kac-Weyl character formula for $\Eh$, we would like to make a final remark on the Macdonald-Weyl denominator and the theta block associated with $\Eh$, inspired from the recent work of Gritsenko, Skoruppa and Zagier \cite{GSZ19}.
The existence of the affine VOA $(\Eh)_k$ at small levels supports that the intermediate Lie algebra $\Eh$ might be extended to a generalized affine Lie algebra. We denote this conjectural algebra by $\hat{E}_{7+1/2}$. Like affine Lie algebras (see \cite{GSZ19}), the formal denominator 
$$
e^{\hat{\rho}}\prod_{\alpha}(1-e^{-\alpha})^{\mathrm{mult}(\alpha)}
$$
may induce a theta block of type
\begin{equation}
\vartheta_{\Eh}(\tau,\mathfrak{z}) =  \eta(\tau)^8 \prod_{r} \frac{\vartheta(\tau,\latt{r,\mathfrak{z}}_{E_7})}{\eta(\tau)}, \quad \tau\in \HH, \; \mathfrak{z}\in E_7\otimes\CC, 
\end{equation}
where $\hat{\rho}$ is the Weyl vector of $\hat{E}_{7+1/2}$,  $\alpha$ takes over all positive roots of $\hat{E}_{7+1/2}$, $r$ runs over all the 
$91$ positive roots of $\Eh$, $\eta$ is the Dedekind eta-function
$$
\eta(\tau) = q^{1/24}\prod_{n=1}^\infty(1-q^n), \quad q=e^{2\pi i\tau},
$$
and $\vartheta$ is the odd Jacobi theta function
$$
\vartheta(\tau,z) = q^{1/8}\zeta^{1/2}\prod_{n=1}^\infty(1-q^n)(1-q^n\zeta)(1-q^{n-1}\zeta^{-1}), \quad z\in\CC,\; \zeta=e^{2\pi iz}.
$$
It is clear that $\vartheta_{\Eh}$ defines a \textit{weak} Jacobi form of weight $4$ and index $24$ for $E_7$. We conjecture that $\vartheta_{\Eh}$ is further a \textit{holomorphic} Jacobi form, that is, it is also holomorphic at infinity (i.e. its Fourier expansion is of type \cite[Equation (26)]{GSZ19}). Such a theta block is exceptional, because it is of critical weight and may be expressed as a $\CC$-linear combination of pullbacks of Jacobi theta functions associated with some positive definite lattices of rank $8$ (see \cite[Equation (36)]{GSZ19}). This might be helpful to determine the Fourier expansion of $\vartheta_{\Eh}$, and further find the infinite sum part of the denominator identity for $\hat{E}_{7+1/2}$. 

Let us consider the more general theta blocks
\begin{equation}
\Theta_d(\tau,\mathfrak{z}) :=  \eta(\tau)^d \prod_{r} \frac{\vartheta(\tau,\latt{r,\mathfrak{z}}_{E_7})}{\eta(\tau)}, \quad d\in \ZZ. 
\end{equation}
From \cite{Wan23} we know that the holomorphic theta blocks of singular weights have a one-to-one correspondence with affine Lie algebras. Therefore, if $\Theta_d$ is holomorphic at infinity then $d\geq 8$. The conjecture above yields that $d=8$ is the minimal number such that $\Theta_d$ defines a holomorphic Jacobi form. The number $8$ may come from the multiplicity of the imaginary roots of $\hat{E}_{7+1/2}$.

\section*{Appendix I}\label{app:I}
As we reviewed, the level 1 affine characters of the Deligne-Cvitanovi\'c exceptional series satisfy an uniform 2nd order MLDE \eqref{eq:level1mlde}. 
In this appendix we present the uniform 6th order MLDE for the level 2 affine characters of the Deligne-Cvitanovi\'c exceptional series. 

Each conformal weight for the Deligne-Cvitanovi\'c exceptional series including $\Eh$ at level $2$ equals one value of the six rational functions 
\begin{align}\label{DC2cw}
0, \frac{h(h+1)}{(h+2)(h+6)}, \frac{h}{h+2}, \frac{5h(h+3)}{3(h+2)(h+6)},\frac{5h-6}{3(h+2)} ,\frac{13h^2-6h+216}{6(h+2)(h+6)} .
\end{align}
Here we use $h$ to represent the dual Coxeter number $h^\vee$ to shorten the expression. It is quite nontrivial that the above six conformal weights always result in an 6th order holomorphic MLDE, i.e., index $l=0$, for arbitrary $h^\vee$.  These conformal weights and the vacuum character $q^1$-term coefficient $\dim(\mathfrak{g})$ enable us to uniquely fix the six coefficients of 6th order MLDE as
\begin{align}
    [{ D}^{6}+\mu_1E_4{ D}^{4}+\mu_2E_6{ D}^{3}+\mu_3E_4^2{D}^{2}+\mu_4E_4E_6{ D}+(\mu_5E_4^3+\mu_6E_6^2)]\chi=0,
\end{align}
where
    \begin{align}
\mu_1&= \frac{-43 h^4+41 h^3-1441 h^2+9168 h-22932}{36 (h+2)^2 (h+6)^2},\\
 \mu_2&=\frac{249 h^6+2778 h^5+468 h^4+80080 h^3-370368 h^2+950400 h-945216}{216 (h+2)^3 (h+6)^3} ,\\
 \mu_3&= -\frac{1}{1296 (h+2)^4 (h+6)^4}(689 h^8+19063 h^7+96981 h^6+292042 h^5+2264117 h^4\\ \nonumber
&\phantom{=}\ -7550088 h^3+11641320 h^2-14000256 h+6413904) ,\\
 \mu_4&= \frac{1}{7776 (h+2)^5 (h+6)^4}(1000 h^9+45175 h^8+327973 h^7+463584 h^6+9223896 h^5\\ \nonumber
&\phantom{=}\ -14947800 h^4-4476960 h^3+1902528 h^2+5500224 h-31104), \\
 \mu_5&= -\frac{(h+1)(5h-6)}{15552 (h+2)^6 (h+6)^6}(4025 h^9+11390 h^8+570314 h^7+208384 h^6+11337365 h^5\\ \nonumber 
 &\phantom{=}\ +6562770 h^4-2328840 h^3-47884176 h^2-56934576 h-48250080),\\
\mu_6&=-\frac{(h+1)(5h-6)}{5832 (h+2)^6 (h+6)^6} (25 h^{10}-25 h^9+21079 h^8-21355 h^7+887152 h^6-1084 h^5\\ \nonumber 
 &\phantom{=}\ +7082880 h^4-3294576 h^3+3451680 h^2+16708032 h+17698176).
\end{align}

Similarly, for affine VOA associated with Deligne-Cvitanovi\'c exceptional series at level three, we find the 12 conformal weights can be written as the rational functions of $h^\vee$ as
\begin{align}\label{DC3cw}
&0, \frac{h(h+1)}{(h+3)(h+6)}, \frac{h}{h+3}, \frac{5h(h+3)}{3(h+3)(h+6)},\frac{5h-6}{3(h+3)} ,\frac{13h^2-6h+216}{6(h+3)(h+6)},\\ 
&\frac{2(h+1)}{(h+6)},\frac{2h}{(h+3)},\frac{5h(h+4)}{2(h+3)(h+6)},\frac{5h-6}{2(h+3)},\frac{h(8h+33)}{3(h+3)(h+6)},\frac{7h^2-12h+216}{2(h+3)(h+6)}.
\end{align}
We notice that the above 12 conformal weights always result in a 12th order MLDE with index $l=20$ for arbitrary $h^\vee$.

\bigskip

\noindent
\textbf{Acknowledgements}
KS would like to thank Tomoyuki Arakawa and Kazuya Kawasetsu for crucial discussions at RIMS. The authors also thank Thomas Creutzig, Zhihao Duan, Ken Kikuchi, Hee-Cheol Kim,  Du Pei, Siddhartha Sahi, Xin Wang, Yi-Nan Wang, Wenbin Yan and Di Yang for useful discussions. KL are supported by KIAS Grants PG006904 and by the National Research Foundation of Korea Grant funded by the Korea government (MSIT) (No.2017R1D1A1B06034369). The majority work of KS was done at KIAS under the support of Grant QP081002. KS would also like to thank the hospitality of MPIM Bonn where this work is finished. Some results have been presented by KS at KIAS, USTC, Tsinghua and in Osaka during the workshop ``Quantum Field Theories and Representation Theory". KL would like to thank the hospitality of Pollica Physics Center where many physicists and mathematicians got together to exchange ideas freely.


\bibliographystyle{plainnat}
\bibliofont
\bibliography{refs}

\end{document}